\documentclass{article}

\usepackage[top=1in, bottom=.75in, left=0.75in, right=0.75in]{geometry}

\usepackage{graphicx}
\usepackage{amsfonts}
\usepackage{amssymb}
\usepackage{amsmath}
\usepackage{mathtools}
\usepackage{subcaption}
\usepackage{xcolor}
\usepackage{url}

\usepackage{setspace}
\makeatletter
\renewcommand{\maketitle}{
  \begin{center}
    {\huge \@title \par}
    \vskip 0.5em
    {\small \@author \par}
    \vskip 0.3em
    {\small \@date \par}
  \end{center}
  \vskip 2em
}
\makeatother
\usepackage[symbol]{footmisc}
\title{A continuum, computational study of morphogenesis in lithium intermetallic interfaces}
\author{
  ~Mostafa Faghih Shojaei$^{a}$,
   ~Rahul Gulati$^{a}$, 
  ~Krishna Garikipati$^{a}$\footnote{e-mail: krishna.garikipati@usc.edu}
  \\[3mm]
   $^a$ \textit{Department of Aerospace \& Mechanical Engineering, University of Southern California}}

\begin{document}
\maketitle

\begin{abstract}
The design of solid state batteries with lithium anodes is attracting  attention for the prospect of high capacity and improved safety over liquid electrolyte systems. The nature of  transport  with lithium as the current carrier has as a consequence the accretion or stripping away of the anode with every charge-discharge cycle. While this poses challenges from the growth of protrusions (dendrites) to loss of contact, there lurks an opportunity: Morphogenesis at the anode-electrolyte interface layer can be studied, and may ultimately be controlled as a factor in solid state battery design. The accessible interface morphologies, the dynamic paths to them, and mechanisms to control them expand considerably if lithium alloys are introduced in the anode. The thermodynamics and kinetics of  lithium intermetallics present principled approaches for morphogenic interface design. In this communication we adopt a computational approach to such an exploration. With phase field models that are parameterized by a combination of first principles atomistic calculations and experiments, we  present phenomenological studies of two lithium intermetallics: Li-Mg and Li-Zn. An array of parametric investigations follows on the influence of kinetics, charge-discharge rate, cycling, transport mechanisms and grain structure. The emphasis across these computations is on the dynamic morphogenesis of the intermetallic interface. Specifically, the plating, segregation and smooth distribution of Li, Mg and Zn, the growth and disappearance of voids, evolution of solid electrolyte-anode contact area, and grain boundary structure are investigated. The computational platform is a  framework for future studies of morphogenic electrolyte-anode interfaces with more extensive inputs from first principles atomistics and experiments.
\end{abstract}

\section{Introduction}

Lithium metal has been widely used in batteries for its high electrochemical potential. Lithium-ion systems constitute an important class of rechargeable batteries with widespread applications in everyday appliances. However, lithium-ion batteries with liquid electrolytes  pose serious safety concerns. The advent of fully solid-state batteries with solid electrolytes has mitigated this safety hazard. Besides eradicating safety concerns, solid state batteries also have the potential to store  energy at higher densities and block lithium dendrites. Perhaps the most widely recognized solid electrolyte is LLZO for its high ionic conductivity and excellent chemical and electrochemical stability when in contact with lithium metal \cite{wang2020garnet, murugan2007fast}. 

However, morphological instabilities at the lithium metal-electrolyte interface significantly constrain the rate capacity of solid state batteries. Dendrite-induced short circuits continue to plague LLZO-based solid-state batteries at high current densities, impeding their practical application. The kinetics of lithium nucleation and growth at the interface of solid state electrolytes and electrodes are influenced by the current density. The critical current density of a battery plays a vital role in delivering the maximum available current density but is limited by factors including lithium kinetics, interface wettability and grain boundaries. Application of external stack pressure can establish the interface contact but can fracture especially brittle ceramic electrolytes such as LLZO. The recent literature has suggested using a lithium alloy anode or an interface solid to avoid the formation of voids at the interface by reinforcing the anode-electrolyte contact \cite{siniscalchi2022relative, krauskopf2019diffusion, koshikawa2019electrochemical, luo2017reducing, fu2017toward, yang2019electron, wan2020mechanical, luo2021growth, lewis2022promise}. We briefly review some of the existing literature on lithium intermetallics, focusing on those aspects that have influenced our study.

It has been proposed that aluminum foil anodes have the potential to revolutionize lithium-ion batteries with their high capacity and low cost. However, owing to their insufficient cycle life, other materials, such as graphite have gained in popularity as anodes. The diffusional lithium trapping in aluminum foil anodes of lithium-ion batteries was investigated by Crowley et al. \cite{crowley2022diffusional}, highlighting its role in capacity degradation. The authors identified as the predominant mechanism, the trapping of lithium ions along the grain boundaries of the $\beta$-Li-Al phase, a process intensified by higher de-lithiation rates and greater areal capacities. By alloying aluminum with lithium and silicon, Crowley et al. demonstrated enhanced first-cycle efficiency and cycle life, due to improvements to the compositional and microstructural optimization of the anodes.

Lithium alloys have delivered enhanced interface morphological stability of solid state batteries, which, it has been speculated, is due to the role of lithium diffusivity in the alloys \cite{krauskopf2019diffusion, wan2020mechanical, lewis2022promise, jin2020solid}. Among  Li alloys, Li-Mg has been widely popular since Mg is lightweight and has a wider solubility range in lithium in the $\beta$-Mg ($\beta$-Li) phase. Siniscalchi et al. carried out an analysis to compare the bulk diffusivity of lithium in the pure metal to that of the lithium alloy \cite{siniscalchi2022relative}. Through their experiments, they revealed that the diffusivity of lithium in Li-Mg is even lower, leading to a higher critical current density for pure lithium metal. Lithium metal electrodes however, suffer from void formation, accompanied by loss of contact while the presence of magnesium, which is not stripped, preserves the contact morphology during cycling. However, the slower bulk lithium diffusivity in the Li-Mg alloy limits the critical current density. Improvement in performance of the Li-Mg anode over pure lithium is not due to the role of Li diffusivity since the pure lithium anode outperforms the alloy for larger current density. However, at a lower stripping current, the capacity extracted from Li-Mg is higher due to its stable interface morphology. 

To avoid void formation at the Li - electrolyte (LLZO) interface and enhance the interface morphological stability, He et al. \cite{he2021cu} studied the effect of a Cu-doped Li-Zn alloy interface layer which guides uniform Li deposition. They demonstrated improved lithium plating, stripping performance, and a high critical current density. The Li-Zn layer promotes faster lithium diffusion. The Cu nanoparticles create an electric field resulting in the homogeneous deposition of lithium and additionally  act as a nucleation site. The authors emphasized the role of the interface in solid state batteries for maximizing their performance. 

The effect of temperature and stack pressure on the performance of lithium and Li-Mg alloy-based solid state batteries was studied by Krauskopf et al. \cite{krauskopf2019diffusion}. The authors observed, as have others, that the  adoption of LLZO as a favored electrolyte has promoted the need for morphological stability of lithium metal anodes. They found  the lithium depletion time of Li-Mg to be higher than that of the bare lithium electrode. The authors also inferred that temperature has a strong effect on lithium diffusion properties and therefore the morphological stability since higher lithium depletion is observed at lower temperatures.

Thus, the study of improving the current charge density while maintaining the morphological integrity of the anode-electrolyte interface has gained traction due to its potential to enhance the performance of solid state batteries. While many experimental insights have been reported in the literature, such as those outlined above, the fundamental operating principles of alloys and alloy interlayers are still not well understood \cite{dai2018interface}. The state of the science leaves much room to assemble a framework for a fundamental understanding of morphogenesis of these interfaces.  Computations at the continuum scale offer opportunities to test mechanisms, translate first-principles knowledge  into parametric studies, and validate hypotheses against experiments. However, to the best of our knowledge, continuum computational studies of the morphogenesis of the anode interface at the continuum scale have not been undertaken in a systematic manner. In this work,  we study this problem of morphogenesis of the lithium intermetallic interface and present a computational framework capable of supplementing the experimental findings. Drawing from the above literature, we focus on lithium-magnesium and lithium-zinc intermetallics, in which we study the influence of kinetics, charge-discharge rates, cycling, transport mechanisms and grain structure. To model  morphogenesis we adopt phase field approaches, previously used by us to describe the multiphysics of  battery materials \cite{shojaei2024bridging,jiang2016multiphysics,rudraraju2016mechanochemical}. We employ  finite element methods to discretize the coupled partial differential equations \cite{wang2018multi,garikipati2017perspective,wang2017intercalation}.

The  governing equations of Cahn-Hilliard dynamics and its numerical treatment are presented in Section \ref{sec:main_formulation}.  A number of case studies of interest are detailed in Section \ref{sec:case_studies} to understand the morphological evolution of the intermetallic solid. We start with the temporal evolution of a generic lithium alloy and subsequently study the charging-discharging dynamics for Li-Mg and Li-Zn. Concluding remarks are presented in Section \ref{sec:conclusion}.

\section{Formulation}\label{sec:main_formulation}

We investigate the morphogenesis of anode interfaces as binary substitutional alloys containing vacancies. These alloys consist of lithium and another metal, such as magnesium or zinc. 
Let $N_{V}$, $N_{M}$, and $N_{Li}$ represent the number of vacant lattice sites, metal (Mg/Zn) lattice sites, and lithium lattice sites, respectively, within a representative volume $V_c$ of the substitutional alloy, and let $S$ denote their total. We define $\Omega$ as the volume per crystal site, given by $V_{c}/S$. We define $x_1 = N_V/S$ as the vacancy composition and $x_2 = N_{M}/S$ as the composition of the metal. By the conservation of lattice sites, $x_3 = N_{Li}/S = 1 - x_1 - x_2$ represents the lithium composition. The alloy's free energy can be expressed as $G(N_{V}, N_{M}, N_{Li})$. By normalizing this by the number of crystal sites $S$, we obtain a free energy per crystal site denoted by $\tilde{g}(x_1, x_2)$.

We use phase field theory to model the evolution of microstructure and study its morphogenesis. In particular, we use the Cahn-Hilliard equation to describe the dynamics of conserved quantities $x_1$ and $x_2$ during their evolution and phase changes. The total free energy of the microstructure, excluding the elastic effects, is written as follows:
\begin{equation}\label{eq:total_energy}
f(x_1,x_2) = \int_V \left(g(x_1,x_2) + \frac{1}{2}\kappa_1|\nabla x_1|^2 + \frac{1}{2}\kappa_2|\nabla x_2|^2 \right)\mathrm{d}v,
\end{equation}
where the energy density per volume $g(x_1,x_2) = {\tilde{g}(x_1,x_2)}/{\Omega} = {G(N_V, N_M, N_{Li})}/{V_c}$ and $V$ denotes the material volume, $|\nabla x_1|^2$ and $|\nabla x_2|^2$ are the dot products of the composition gradients, and $\kappa_1$ and $\kappa_2$ are the gradient parameters related with the energy of the interphase interfaces. In the context of phase field theory, $x_1$ and $x_2$ are treated as field variables that are functions of time $t$ and spatial coordinates and have a range of $[0,1]$. The Cahn-Hilliard equations are
\begin{equation}\label{eq:CH}
\frac{\partial x_i}{\partial t} = -  \Omega \nabla \cdot (-L_i \nabla \mu_i), \quad i=1,2,
\end{equation}
where $\nabla \cdot$  denotes the divergence operator, $L_i$ are the mobility coefficients (kinetic-transport coefficients \cite{van2010vacancy}) with  units of $\text{length}^2/(\text{energy} \cdot \text{time})$, and $\mu_i$ are chemical potentials obtained as variational derivatives of the free energy density functional in Eq. (\ref{eq:total_energy}) resulting in
\begin{equation}\label{eq:chem}
\mu_i =\frac{\partial g}{\partial x_i} -\kappa_i\nabla^2x_i, \quad i=1,2,
\end{equation}
where $\nabla^2$ is the Laplacian operator in  Cartesian coordinates. We solve Eqs. \ref{eq:CH} and \ref{eq:chem} simultaneously by first converting them to weak form as follows: 
\begin{equation}\label{eq:weakform}
\begin{aligned}
    0 &= \int_V \left[ w_{x_1}\frac{\partial x_1}{\partial t} + M_1\nabla w_{x_1}\cdot\nabla{\mu}_1 \right]\mathrm{d}v + \int_{\partial V} w_{x_1} f_{n_1}\mathrm{d}s\\
    0 &= \int_V \left[w_{\mu_1}\left(\mu_1 - \frac{\partial g}{\partial x_1}  \right) - \kappa_1\nabla w_{\mu_1}\cdot\nabla x_1\right]\mathrm{d}v\\  
    0 &= \int_V \left[ w_{x_2}\frac{\partial x_2}{\partial t} + M_2\nabla w_{x_2}\cdot\nabla{\mu}_2\right]\mathrm{d}v + \int_{\partial V}  w_{x_2} f_{n_2}\mathrm{d}s\\
    0 &= \int_V \left[w_{\mu_2}\left( \mu_2 - \frac{\partial g}{\partial x_2}  \right) - \kappa_2\nabla w_{\mu_2}\cdot\nabla x_1\right]\mathrm{d}v,\\  
\end{aligned}
\end{equation}
where $w_{x_i}$ and $w_{\mu_i}$, $i = 1,2$ represent arbitrary $C^0$-regular scalar-valued functions defined on the domain $V$,  $M_i$ = $\Omega L_i$ and $f_{n_i}= \Omega j_{n_i}$ is a scaled boundary outflux. Also, the boundary conditions for $x_i$ and $\mu_i$ used to derive Eq. (\ref{eq:weakform}) are given by:
\begin{subequations}
\begin{equation}\label{eq:bcx}
\nabla x_i  \cdot  \mathbf{n} = 0, \quad i = 1,2
\end{equation}
\begin{equation}\label{eq:mubc}
-L_i \nabla \mu_i \cdot \mathbf{n} = j_{n_i}, i = 1,2,
\end{equation}
\end{subequations}
where $\mathbf{n}$ is the outward unit normal vector to the boundary surface of the material, denoted by $\partial V$. Here, $j_{n_1}$ represents the boundary outflux for vacancies, and $j_{n_2}$ denotes the boundary outflux for metal ions. By conservation, $j_{n_1} + j_{n_2} + j_{n_3} = 0$, where $j_{n_3}$ is the lithium ion flux. Assuming $j_{n_2} = 0$, we reinterpret $j_{n_1} = -j_{n_3}$ as the boundary influx for lithium ions in Eq. (\ref{eq:weakform}). Note that Eq. (\ref{eq:bcx}) is a higher-order Dirichlet boundary condition arising from the derivation   of the chemical potentials as variational derivatives  while enforcing equilibrium at the boundaries, while  Eq. (\ref{eq:mubc}) represents a Neumann boundary condition.

In this work, we assume uncoupled diffusivity, where the mutual effects between diffusion of lithium ions and metal (magnesium or zinc) ions in the binary intermetallic alloy are considered negligible. Let $D_i$, with units of $\text{length}^2/\text{time}$, represent the uncoupled diffusivity coefficient for ions corresponding to the composition $x_i$. The diffusivity coefficients are related to the mobility coefficients $L_i$ as follows \cite{van2010vacancy}:
\begin{equation}\label{eq:D_L}
    D_i = \tilde{L}_i \theta_i, \quad \tilde{L}_i= \Omega k_{B}T L_i, 
\end{equation}
where $k_{B}$ denotes the Boltzmann constant, $T$ stands for temperature, and $\theta_i$ is the thermodynamic factor obtained from the Hessian of the free energy $g$ according to:
\begin{equation}\label{eq:theta}
    \theta_i = \frac{1}{k_BT}H_i, \quad H_i = \frac{\partial^2 g}{\partial x_i^2}. 
\end{equation}
Combining  Eqs. (\ref{eq:D_L}) and (\ref{eq:theta}), we obtain a direct expression for $D_i$:
\begin{equation}\label{eq:mobility}
    D_i = \Omega L_i H_i = M_i H_i.
\end{equation}
This relation is instrumental in arriving at physically informed values for $M_i$ in our phase field weak formulation Eq. (\ref{eq:weakform}) using the available experimental measurements of diffusivity coefficients for lithium and other metals.

\section{Case studies} \label{sec:case_studies}
In this section we explore the morphogenic evolution of an intermetallic layer at the interface of a lithium anode and solid electrolyte using the phase field model introduced in the preceding section. Starting from an idle battery, we consider an intermetallic layer with dimensions 80 $\mu \text{m}$ x 80 $\mu \text{m}$ x 8 $\mu \text{m}$.  From an initial porous structure, the respective compositions of Li, metal and vacancies evolve as described by the mass-conserving Cahn-Hilliard dynamics. A Li-Mg free energy function informed by  Density Functional Theory (DFT) studies \cite{behara2024fundamental} is  constructed to study the charge-discharge kinetics. During the charging process, there is lithiation i.e. lithium flux into the intermetallic region on the anode side of the solid electrolyte-anode interface. The dynamics during delithiation, i.e. the discharging process, is also studied using the phase field simulations. The numerical simulations informed by DFT calculations also present  insight to the Li-Zn intermetallic. Finally, we consider  morphogenic evolution at grain boundaries which act as a source of vacancies. The various case studies are addressed in the following subsections. For each of the cases, the free energy function is presented along with the simulation outcomes.  \\

\subsection{A binary intermetallic interface layer}
In this section we develop a phenomenological free energy density surface for Li-Metal alloys with vacancies and use it to model dynamic morphogenesis at the anode interface of a  Li-ion solid-state battery, aiming to demonstrate that the phase field model and the numerical framework accurately resolve the relevant phenomena.

\subsubsection{Free energy of a binary alloy with vacancies}
\begin{figure}[h]
    \centering
    \includegraphics[width=.8\linewidth]{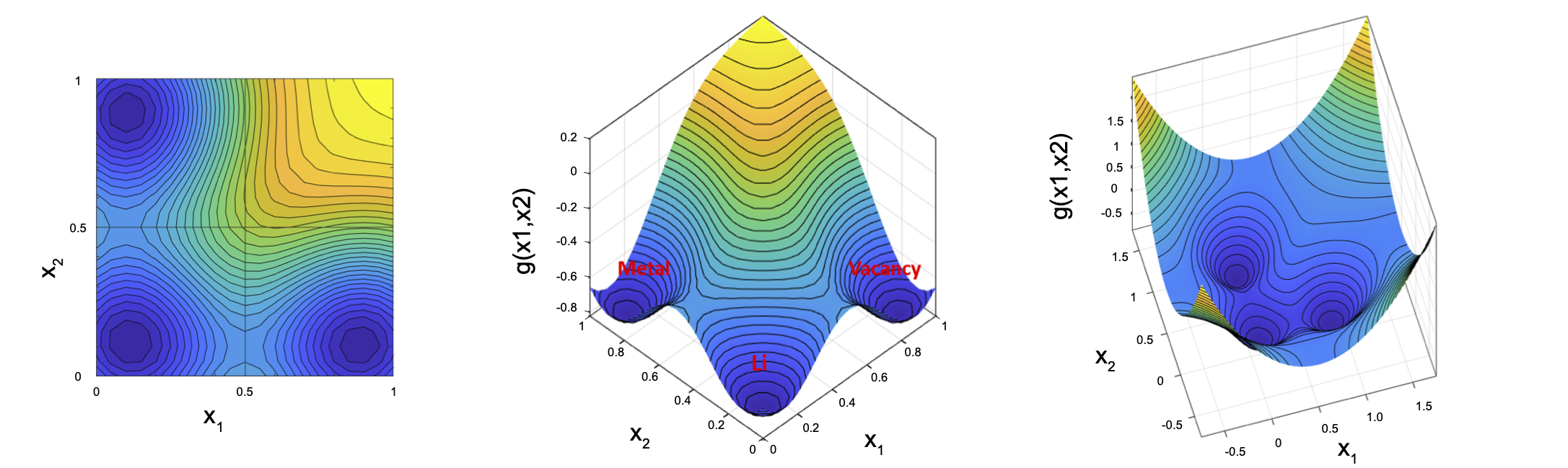}
    \caption{The free energy surface of the Li-metal alloy viewed from different angles, as defined by the expression in Eq. (\ref{eq:Li_metal_enr}). The three minima  representing metal, lithium and vacancy are depicted.
    }
    \label{fig:Li_metal_enr}
\end{figure}

We first propose the following phenomenological free energy surface for Li-metal alloys with vacancies:
\begin{equation}\label{eq:Li_metal_enr}
    \tilde{g}(x_1, x_2) = (x_1 - a)^2 (x_2 - b)^2 + \sum_{i=1}^{4} v_i \exp\left(-r_i \left((x_1 - \bar{x}_i)^2 + (x_2 - \bar{y}_i)^2 \right)\right),
\end{equation}
where $x_1$ and $x_2$ represent the compositions of vacancies and the metal, respectively. The parameters are assigned as follows:: $ a = b = 0.5 $, $ \bar{x} = [0.1, 0.9, 0.1, 0.45] $, $ \bar{y} = [0.1, 0.1, 0.9, 0.45] $, $ v = [-1, -1, -1, 0.2] $, and $ r = [7, 7, 7, 0.5] $. They determine the locations and shapes of four Gaussian  functions added in Eq. (\ref{eq:Li_metal_enr}) to create three  wells on the energy surface, as illustrated in Figure \ref{fig:Li_metal_enr}. Starting with a random distribution of $x_1$ and $x_2$ as the initial configuration of the anode interface, we expect that due to Cahn-Hilliard dynamics of spinodal decomposition and Ostwald ripening, the microstructure  will evolve and stabilize at compositions corresponding to the three energy wells. We call attention to the location of the vacancy well at $(x_1,x_2) = (1,0)$, the lithium well at $(0,0)$ and the metal well at (0,1). The free energy density written as a function of $x_1$ and $x_2$ enforces formation of vacancies on the lithium sub-lattice--with associated phenomenology that we point out in connection with the simulation results below. Density Functional Theory (DFT) studies \cite{behara2024fundamental} informed subsequent versions of the free energy density for Li-vacancy  and Li-metal binary systems.

\subsubsection{A baseline model for formation of a Li-metal interface layer with vacancies}
\label{sec:baselinemodel}
\begin{figure}[h]
    \centering
    \includegraphics[width=.65\linewidth]{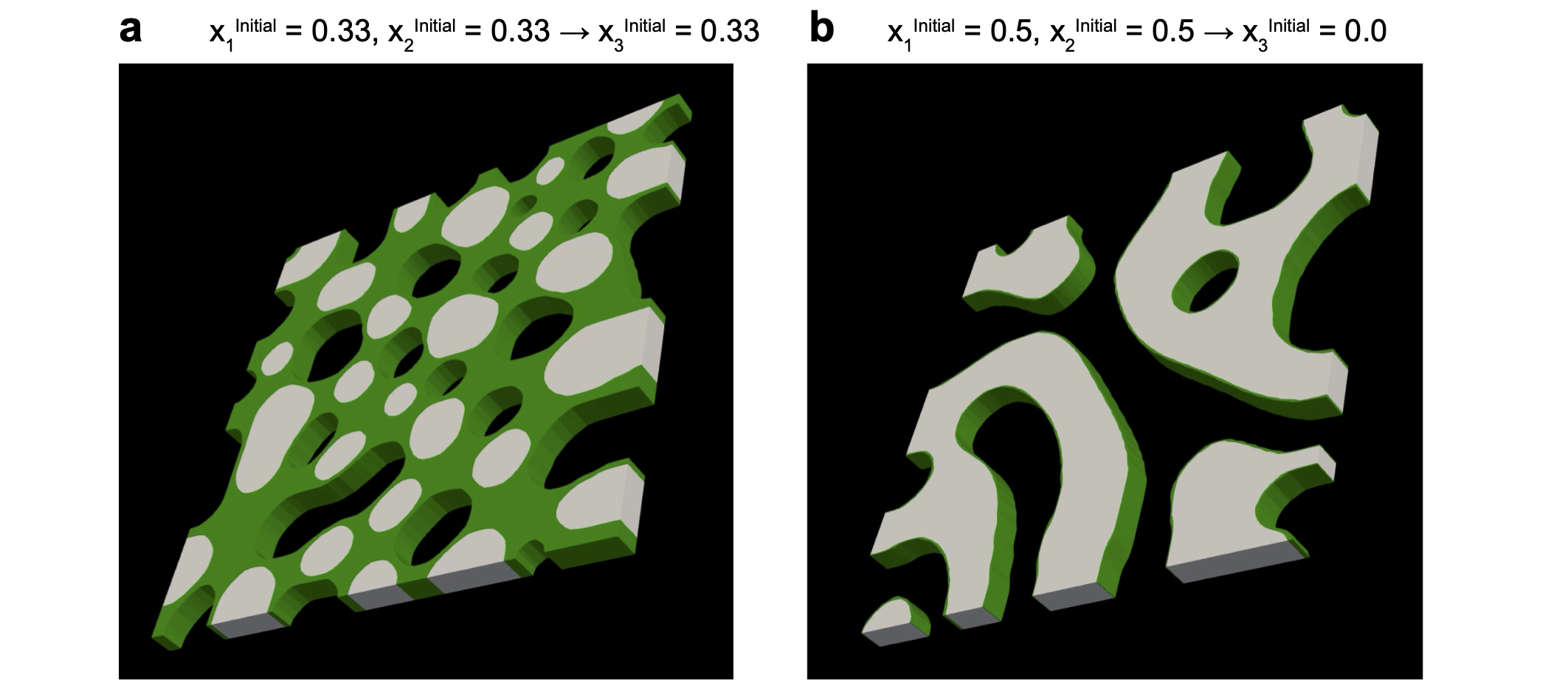}
    \caption{\footnotesize A lithium-metal intermetallic interface possessing the desired level of porosity can be established by choosing different initial conditions on the respective compositions. Lithium is present in (a) but absent in (b) as would be the case after complete discharge of the anode. The green color represents lithium while the white region depicts metal. }
    \label{fig:ostwaldRipening}
\end{figure}

We first study, as a baseline case for the morphogenesis of the intermetallic interface, an initial and boundary value problem with no lithium exchange with the surroundings. From the initial condition, we can establish a porous intermetallic layer with the desired porosity as demonstrated in Figure \ref{fig:ostwaldRipening}. The compositions of Li, metal, and vacancy are equal initially in case (a) as indicated in the figure. This case is representative of a scenario where the battery is in an idle state before being charged or discharged. In Figure \ref{fig:ostwaldRipening}b, the initial Li composition is $x_3 = 0$ and  the evolution consists of a redistribution of metal and vacancies since there is no flux exchange in this case. This case models the anode after being completely discharged. The free energy function in Eq. (\ref{eq:Li_metal_enr}) models vacancies  on the lithium sub-lattice. As is apparent from the figures, this prevents the formation of metal-void interfaces in favor of voids in the lithium regions.

\begin{figure}[h]
    \centering
    \includegraphics[width=.75\linewidth]{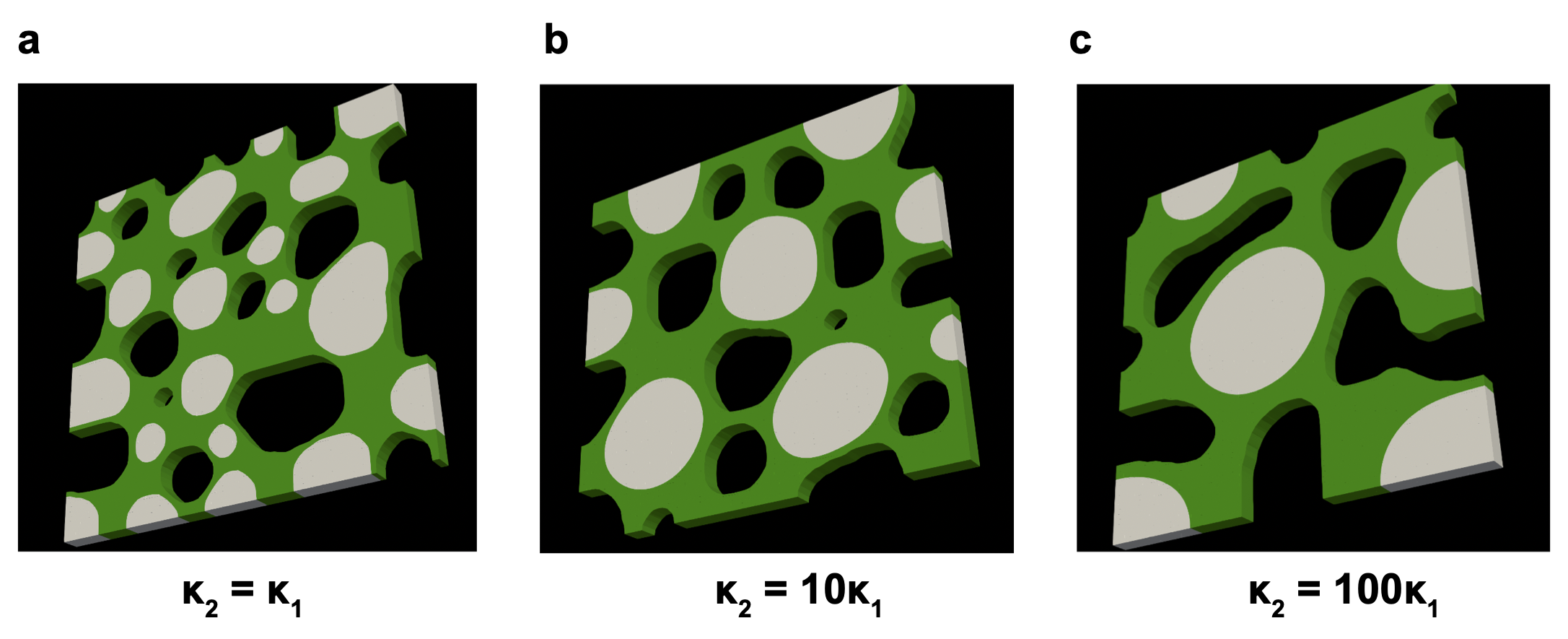}
    \caption{\footnotesize The gradient length scale parameter $\kappa_2$ drives morphogenic structure formation by penalizing  the formation of lithium-metal interfaces, leading to larger metal particles for higher values of $\kappa_2$.}
    \label{fig:ostwaldRipening2}
\end{figure}

Next, the role of the gradient length scale parameters $\kappa_1$, $\kappa_2$ on the morphogenic evolution is established.  Figure \ref{fig:ostwaldRipening2} shows that higher values of the gradient length scale parameter $\kappa_2$, lead to larger metal particles and smaller lithium -metal interface length. As would be expected from phase field models, this follows from the role of $\kappa_2$ as a penalty on the gradient of the  composition value, $\nabla x_2$, in Eq. (\ref{eq:Li_metal_enr}). The cost of the lithium-metal interface, with a composition gradient, $\nabla x_2$, is higher than the lithium-void interface, leading to smaller total lithium-metal interface length and larger, more circular metal particles with smoother boundaries.

\subsection{Morphogenesis in Li-Mg interface layers}
Among lithium intermetallics, we begin with Li-Mg. Recall from the Introduction, that the wide range of solubility of the $\beta$-Mg (sometimes referred to as  $\beta$ Li) at room temperature allows deep stripping of lithium leaving behind magnesium to maintain electrical contact in interface layers. We  study the temporal evolution of the Li-Mg intermetallic layer. As in Section \ref{sec:baselinemodel} we begin by constructing the free energy density function followed by a series of initial and boundary value problems of interface layer morphogenesis.

\subsubsection{DFT-informed free energy of Li-Mg alloy with vacancies}

\begin{figure}[h]
    \centering
    \includegraphics[width=0.7\linewidth]{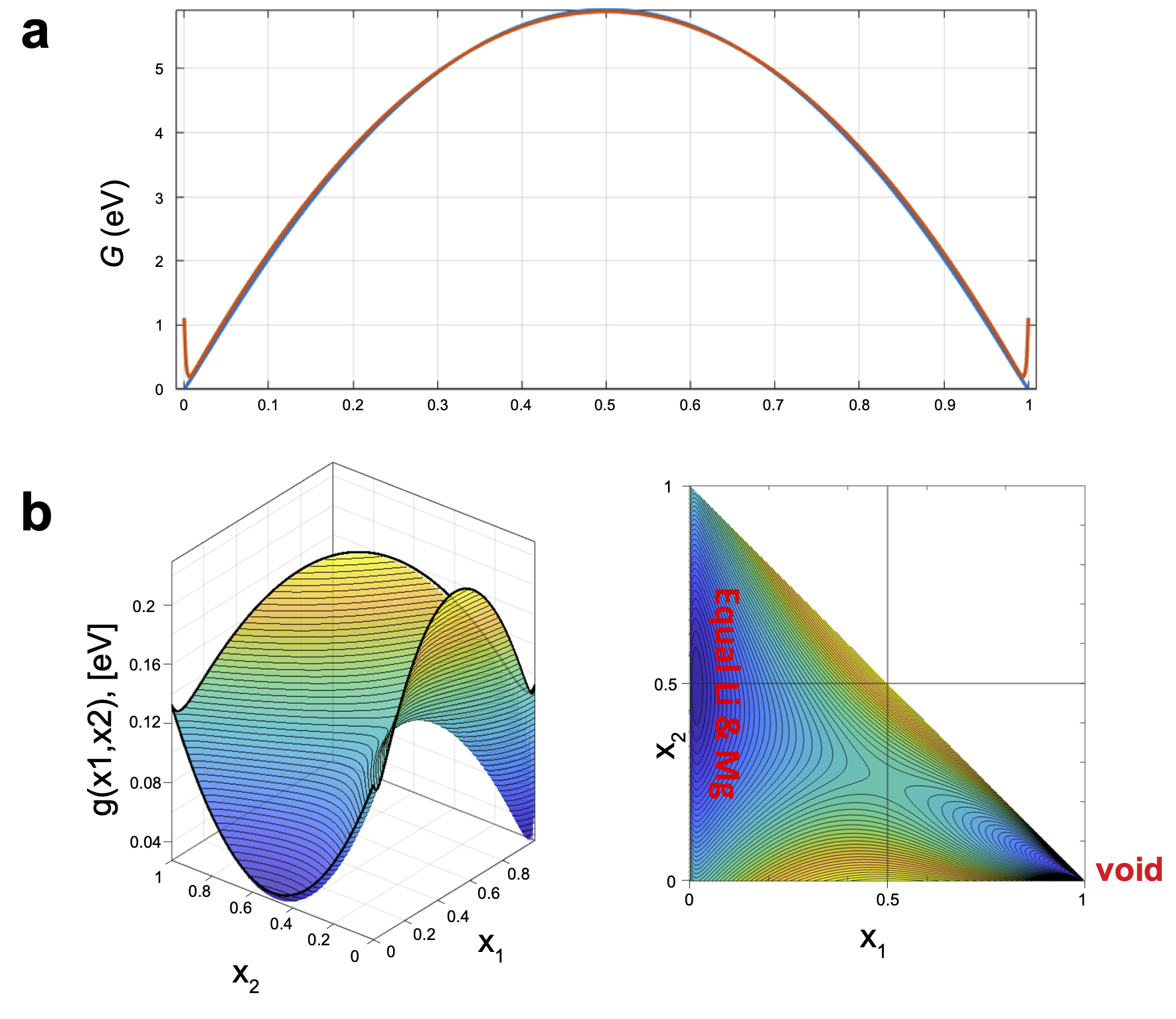}
    \caption{\footnotesize Plots of the Li-vacancy free energy function (Eq. (\ref{eq:Li_va_enr0}) in blue, and its numerically-stabilized approximation (Eq. (\ref{eq:Li_va_enr0_mod}) in red, are shown in a). Surface plots of the two-dimensional free energy function for the Li-Mg alloy containing vacancies (Eq. (\ref{eq:Li_Mg_enr}) are displayed in b). Note that $x_1+x_2 \le 1$. The plot on the right emphasizes the constraint $x_1+x_2 \le 1$, implying $x_3 \ge 0$. }
    \label{fig:Li_Mg_vac_enr}
\end{figure}

\begin{figure}[h]
    \centering
    \includegraphics[width=0.65\linewidth]{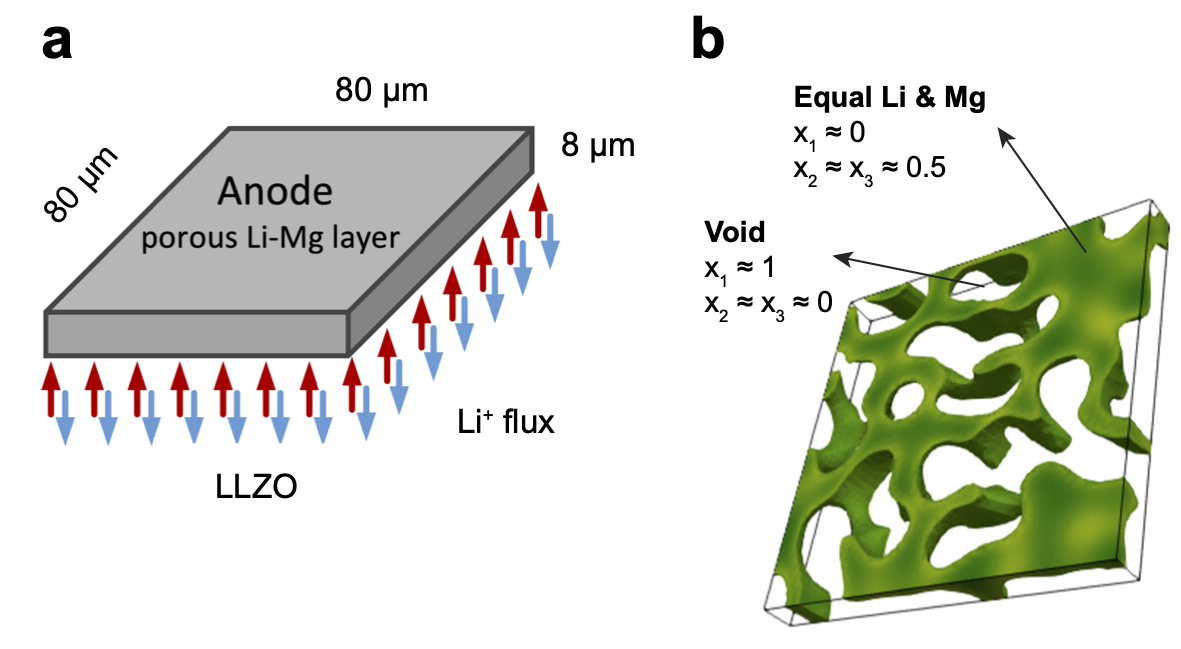}
    \caption{(a) Intermetallic layer modelled for the interface study. 
    (b) Initial configuration of the Li-Mg porous interface layer.}
    \label{fig:free-energy}
\end{figure}

Guided by DFT data \cite{behara2024fundamental}, we develop the following free energy function for a lithium crystal with vacancies, in terms of the vacancy composition:
\begin{equation}\label{eq:Li_va_enr0}
    \bar{g}_1(x_1) = (a+b)x_1 + cx_1^2 + k_BT \left( x\ln(x_1) + (1-x_1)\ln(1-x_1) \right).
\end{equation}

The term 
$k_BT ( x\ln(x_1) + (1-x_1)\ln(1-x_1)) $
accounts for the configurational entropy of mixing for the vacancies on the Li  sub-lattice, and the parameters for the quadratic polynomial are calibrated by running several statistical mechanics computations, underpinned by a large set of high-precision DFT computations \cite{behara2024fundamental}. The fitted values for the coefficients are as follows: $a = -1.90436 \, \text{eV}$, $b = 2.452463 \, \text{eV}$, and $c = -0.548107 \, \text{eV}$. The Boltzmann constant $k_B$ is $8.617 \times 10^{-5} \, \text{eV/K}$, and the temperature $T$ of the system is $300 \, \text{K}$. 

The energy wells of $\tilde{g}_1(x_1)$, Eq. (\ref{eq:Li_va_enr0}), are positioned at $x_1 = 0.6 \times 10^{-9}$ and $x_1 = 1 - (0.6 \times 10^{-9})$. The proximity of the energy wells to the extremities of the vacancy composition range $[0,1]$ introduces numerical stiffness and potential for early divergence of the computational solutions. To mitigate this numerical stiffness, we introduce an approximation to the function \( \bar{g}_1(x_1) \), defined as:
\begin{equation}\label{eq:Li_va_enr0_mod}
\tilde{g}_1(x_1) = \alpha\left[(a + b)x_1 + c x_1^2 + k_B T \left( \exp(-\beta x_1) + \exp(\beta(x_1 - 1)) \right) \right],
\end{equation}
where $\alpha=0.865$ and $\beta=100$ are calibrated dimensionless parameters optimized for numerical stability. In Figure \ref{fig:Li_Mg_vac_enr}, the energy profiles of the original function $\tilde{g}_1(x_1)$ and the numerically-stabilized approximation $g_1(x_1)$ are illustrated by the blue and red lines, respectively. The modification shifts the energy wells of $g_1(x_1)$ slightly inwards from the edges of the domain, where the original energy wells are located, thus making the function more suitable for numerical simulations while preserving the characteristics of the original DFT-derived energy function.

Given the wide solubility range, the free energy of the Li-Mg alloy in the $\beta$ phase without vacancies can be effectively modeled by a quadratic form with a stable minimum energy state at equal compositions of Li and Mg. Therefore, we write the free energy per crystal site of the Li-Mg alloy containing vacancies as: 
\begin{equation}\label{eq:Li_Mg_enr}
\tilde{g}(x_1,x_2) = \tilde{g}_1(x_1) + p\left(\frac{x_2}{(1-x_1)}-0.5\right)^2 + q,
\end{equation}
where $p=0.4024 \, \text{eV}$ and $q = 0.0101 \, \text{eV}$. We meticulously calibrated these phenomenological parameters by analyzing the free energy response across multiple phase-field simulations.
 Note that, in the absence of vacancies ($x_1=0$), the function $g(x_1=0,x_2)$ simplifies to a quadratic form $p(x_2-0.5)^2 + q$, which attains its minimum at $x_2=0.5$. Similarly, when $x_2=0$, $g(x_1,x_2=0)$ reduces to Eq. (\ref{eq:Li_va_enr0_mod}), up to an additive constant. Figure \ref{fig:Li_Mg_vac_enr} illustrates $g(x_1,x_2)$ and its two energy wells at $(x_1,x_2,x_3) = (0+\varepsilon,0.5,0.5-\varepsilon)$, representing an equimolar Li-Mg phase, and $(x_1,x_2,x_3)=(1-\varepsilon,0,\varepsilon)$ corresponding to voids ($0 < \varepsilon \ll 1$).
 
To determine the baseline values for mobility constants, we used Eq. (\ref{eq:mobility}), calculating the Hessian  of the free energy density function from $g = \tilde{g}(x_1,x_2)$ in Eq. (\ref{eq:Li_Mg_enr}) and utilizing experimental values for diffusivities. For simplification, we assumed that $H_{\text{Li}}$ and $H_{\text{Mg}}$ (diagonal terms of the Hessian) remain constant and equal to their values at $x_1 = 0.001$ and $x_2 = 0.5$. This point is very close to the energy well illustrated in Figures \ref{fig:Li_Mg_vac_enr}b and \ref{fig:Li_Mg_vac_enr}c. 
For diffusivity, we used $D_{\text{Li}} = 3.4958 \times 10^{-10}$ cm$^2$/s and $D_{\text{Mg}} = 4.0 \times 10^{-13}$ cm$^2$/s, which closely align with the experimental values reported in Refs. \cite{siniscalchi2022relative,krauskopf2019diffusion}. These diffusivity values are crucial as they ensure that our simulated diffusion processes are realistic and comparable to observed behaviors. Using these experimentally derived diffusivity values in Eq. (\ref{eq:mobility}), we calculated the corresponding mobilities: $M_{\text{Li}} = 0.04$ and $M_{\text{Mg}} = 0.003$ $\mu$m$^2$/(eV·min). These baseline mobility values were incorporated into our phase field weak formulation to accurately model Li/Mg diffusion within the alloy.

\begin{figure}[h!]
    \centering
    \includegraphics[width=0.8\linewidth]{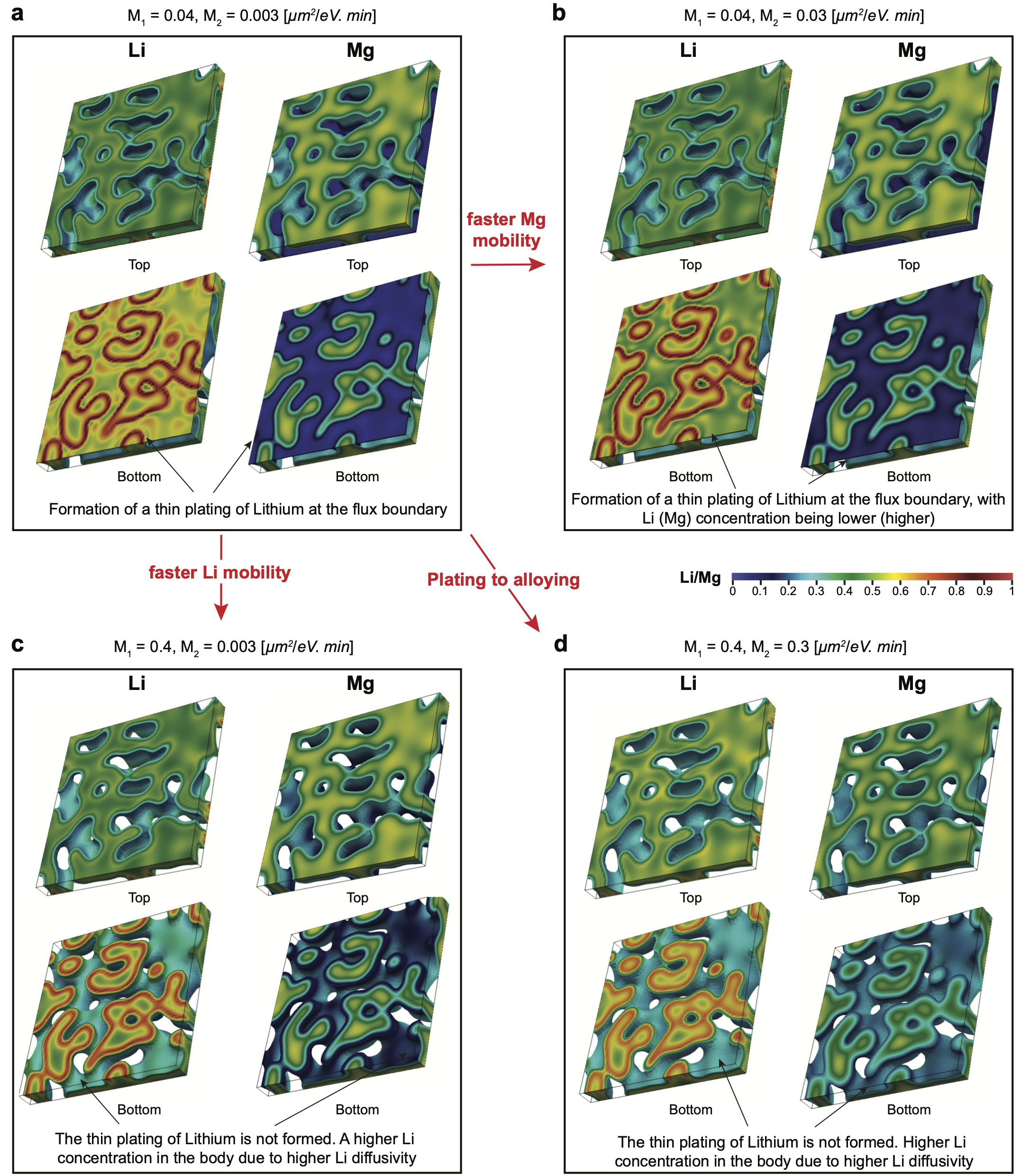}
    \caption{\footnotesize The effect of mobility on morphogenesis of lithium intermetallic solid during charging. (a) depicts the evolution of the top and bottom surface of lithium and magnesium respectively under lithiation. Starting from the same initial condition but with a higher magnesium mobility, higher lithium mobility, a higher lithium and magnesium mobility is depicted in case (b), (c ) and (d) respectively. Slower lithium mobility leads to lithium plating at the contact surface as visible in cases (a) and (b). }
    \label{fig:node-Lithiation-Base-Case}
\end{figure}

\subsubsection{Parametric studies of mobility}
To study the morphogenesis of the Li-Mg intermetallic interface, we consider an initially porous Li-Mg layer generated by spinodal decomposition on the Li-vacancy axis in a ternary Li-Mg-vacancy alloy as depicted in Figure \ref{fig:free-energy} (b). A lithium influx of 1 mA/cm$^2$ is imposed along the bottom surface during charging of the battery. Recall that lithium alloys are important  in solid state batteries for the possibility of enhancing the effective lithium diffusivity to mitigate the formation of voids at the Li (current) influx surface and for thus improving the critical current density of the solid-state battery. The effect of the mobility of lithium and magnesium on the evolution is presented in Figure \ref{fig:node-Lithiation-Base-Case}. The green color represents the equi-compositional Li-Mg alloy ($x_2,x_3 = 0.5$). The top and bottom surfaces showing each of lithium and magnesium are depicted in the figure. Note that the lithium flux exchange is at the bottom surface where we observe a thin plating of lithium at low mobility. On increasing the mobility of lithium, plating is not observed at the contact surface, but transport occurs into the layer, leading to higher lithium composition within the intermetallic structure and void formation at the influx surface. On the other hand, a higher magnesium mobility does cause lithium plating  at the flux boundary, while lower lithium levels are attained within the layer. We thus conclude that slower diffusivity accompanies lithium plating and magnesium trapping during charging. It can be inferred that during discharging at a higher lithium flux rate, a higher lithium mobility is required to mitigate the formation of voids at the contact surface by transporting lithium from within the layer. (Simulations for discharging are presented in the following sections.) As reflected in the figure, during charging, the layer morphogenesis progresses from plating to alloying as Li mobility increases.

\begin{figure}[h!]
    \centering
    \includegraphics[width=1\linewidth]{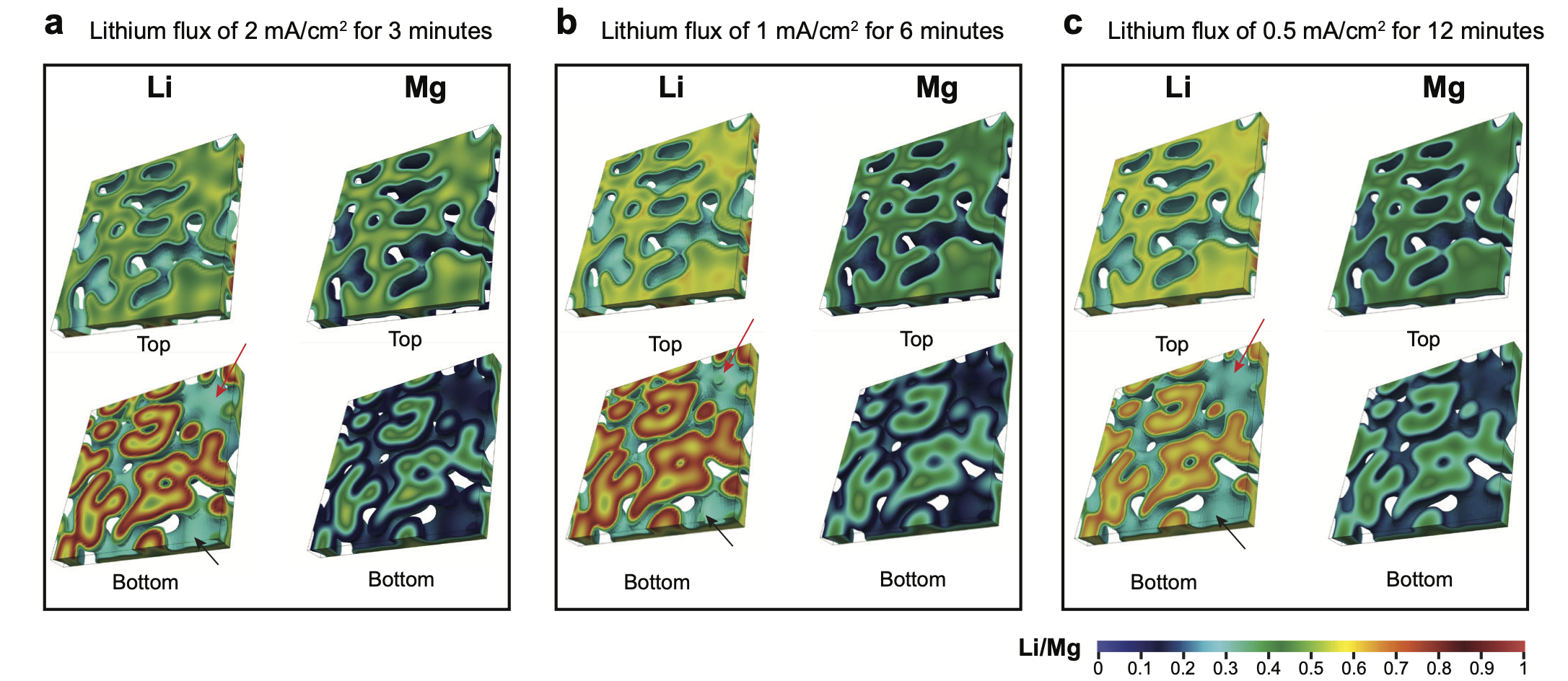}
    \caption{\footnotesize Starting from the same initial condition, (a), (b), (c) illustrate the intermetallic interface after a Li currents (fluxes) of $2 mA/cm^2$ for 3 minutes, $1 mA/cm^2$ for 6 minutes, $0.5 mA/cm^2$ for 12 minutes respectively. As is clear, the net lithium transported is identical in all three cases. Higher lithium fluxes during the charging of the solid-state battery lead to lithium plating at the influx surface. The mobility values for lithium and magnesium in all the cases are taken as $0.04$ $\mu m^2/eV \cdot min$ and $0.003$ $\mu m^2/eV \cdot min$, respectively. 
}
    \label{fig:flux}
\end{figure}

The same effect can be observed from Figure \ref{fig:flux} where the same total amount of lithium is transported into the layer over different times. Due to competition between influx rates and the kinetics, adding the same net lithium at the boundary but at a different rate leads to different microstructures by morphogenesis. We also note that the higher flux applied over a short duration allows for less lithium being transported into the intermetallic structure and therefore results in lithium plating at the contact surface: fast charging leads to lithium plating at the contact surface. This computational exploration of the parameter space can supplement physical experiments and guide the choice of optimum charging rates for the particular intermetallic alloy to avoid lithium plating. Note that lithium plating can be a precursor to dendrites formation.

\begin{figure}[h!]
    \centering
    \includegraphics[width=0.9\linewidth]{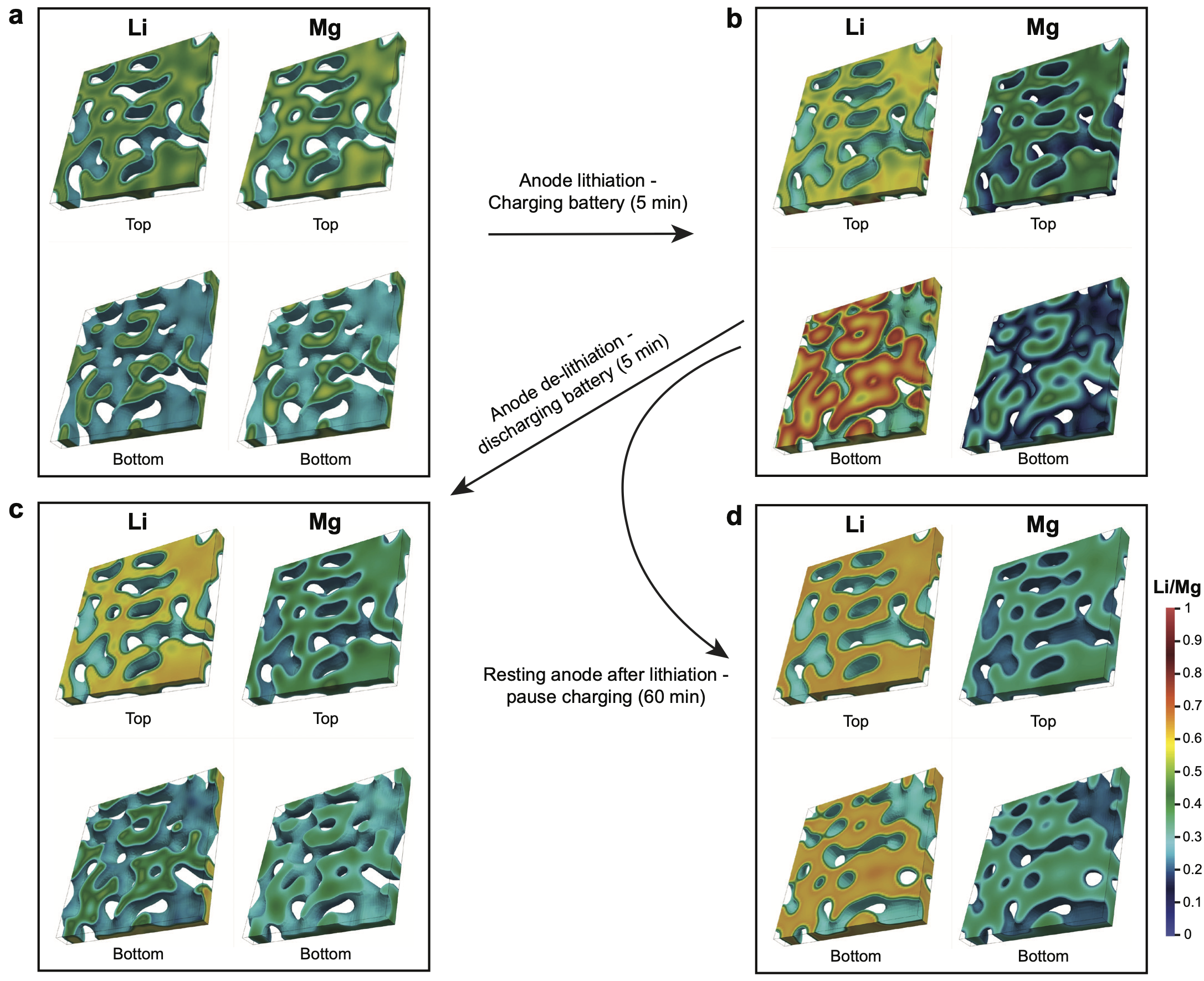}
    \caption{\footnotesize Morphogenesis of an intermetallic layer during a resting period following anode lithiation. (a) represents the initial microstructure of the intermetallic layer, which upon lithiation produces (b). De-lithiation of (b) to remove  lithium equal in mass to that deposited in (a) leads to microstructure (c). Resting the battery in case (b) for an hour leads to (d). Pauses during charge/discharge thus lead to contrasting intermetallic microstructures.}
    \label{fig:resting}
\end{figure}

\subsubsection{Resting period}

Morphogenesis of the intermetallic layer during a resting period is presented in Figure \ref{fig:resting}. We explore the  lack of perfect reversibility during  morphogenesis. The microstructure of the intermetallic solid after charging (Figure \ref{fig:resting}a to \ref{fig:resting}b) and discharging for equal durations evolves to a  microstructure (Figure \ref{fig:resting}c) that does not correspond to the starting configuration (Figure \ref{fig:resting}a). Furthermore, as may be expected, during the resting phase, the microstructure continues to evolve: the lithium composition on the bottom surface decreases and increases on the top surface due to internal transport (Figure \ref{fig:resting}b to \ref{fig:resting}d).  In addition to the potential for extension to several charge-discharge cycles, the detailed dynamics also reveal the characteristics of void formation at the contact surface during discharging (Figure \ref{fig:resting}b to \ref{fig:resting}c) and healing during resting (Figure \ref{fig:resting}b to \ref{fig:resting}d).   

\begin{figure}[h!]
    \centering
    \includegraphics[width=0.9\linewidth]{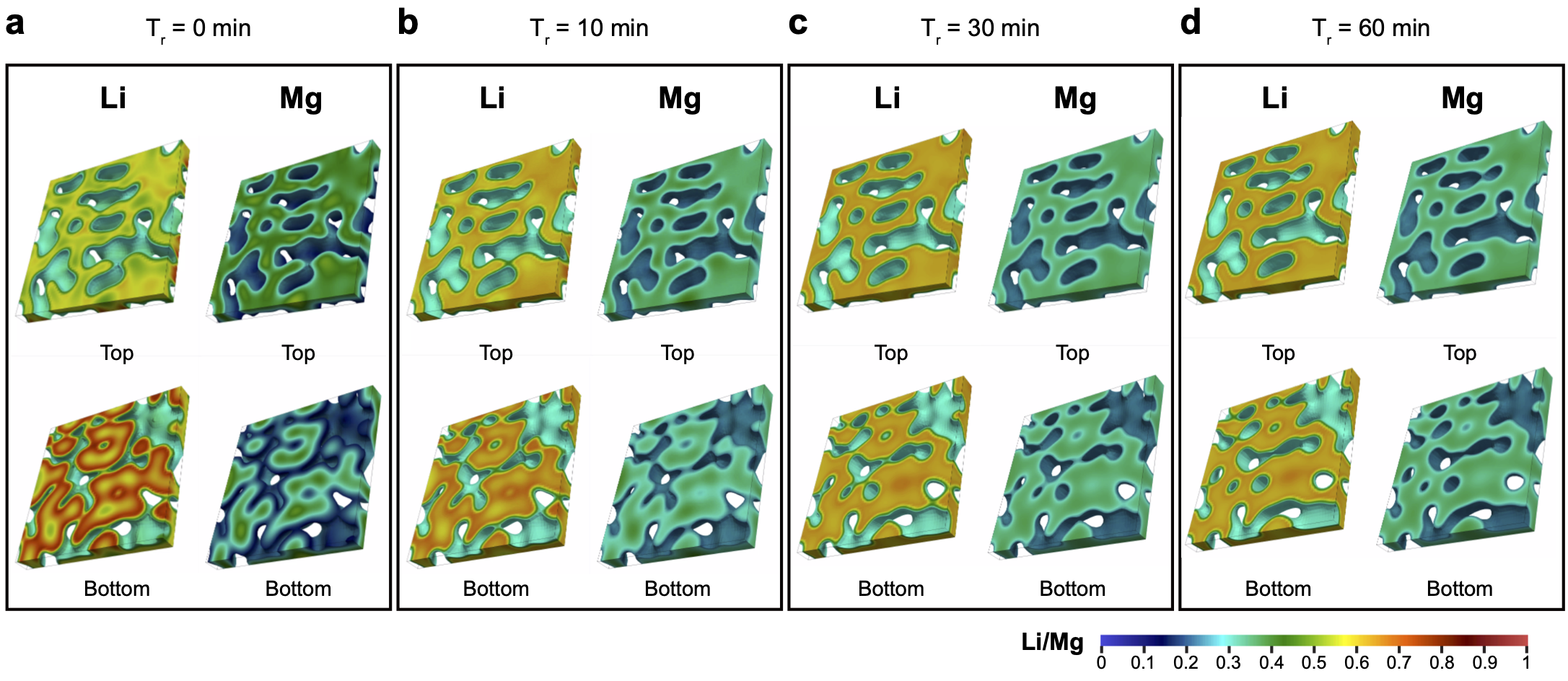}
    \caption{Snapshots of the interface layer at instants during a 60-minute resting period post-lithiation.	During the resting phase, there is no lithium-ion flux at the contact surface. Both, the peak lithium composition at the contact  surface and the void surface area decrease during resting.}
    \label{fig:resting2}
\end{figure}

To study the temporal evolution during the resting period, Figure \ref{fig:resting2} provides  insight to morphogenesis at different time frames. We note that the lithium plating at the contact surface formed due to fast charging dissolves with time and lithium diffuses into the bulk volume even reaching the top surface.  A resting period after  fast charging  of the battery promotes more even distribution of lithium and metal, and can increase the cycle life of the battery.

\begin{figure}[h!]
    \centering
    \includegraphics[width=0.85\linewidth]{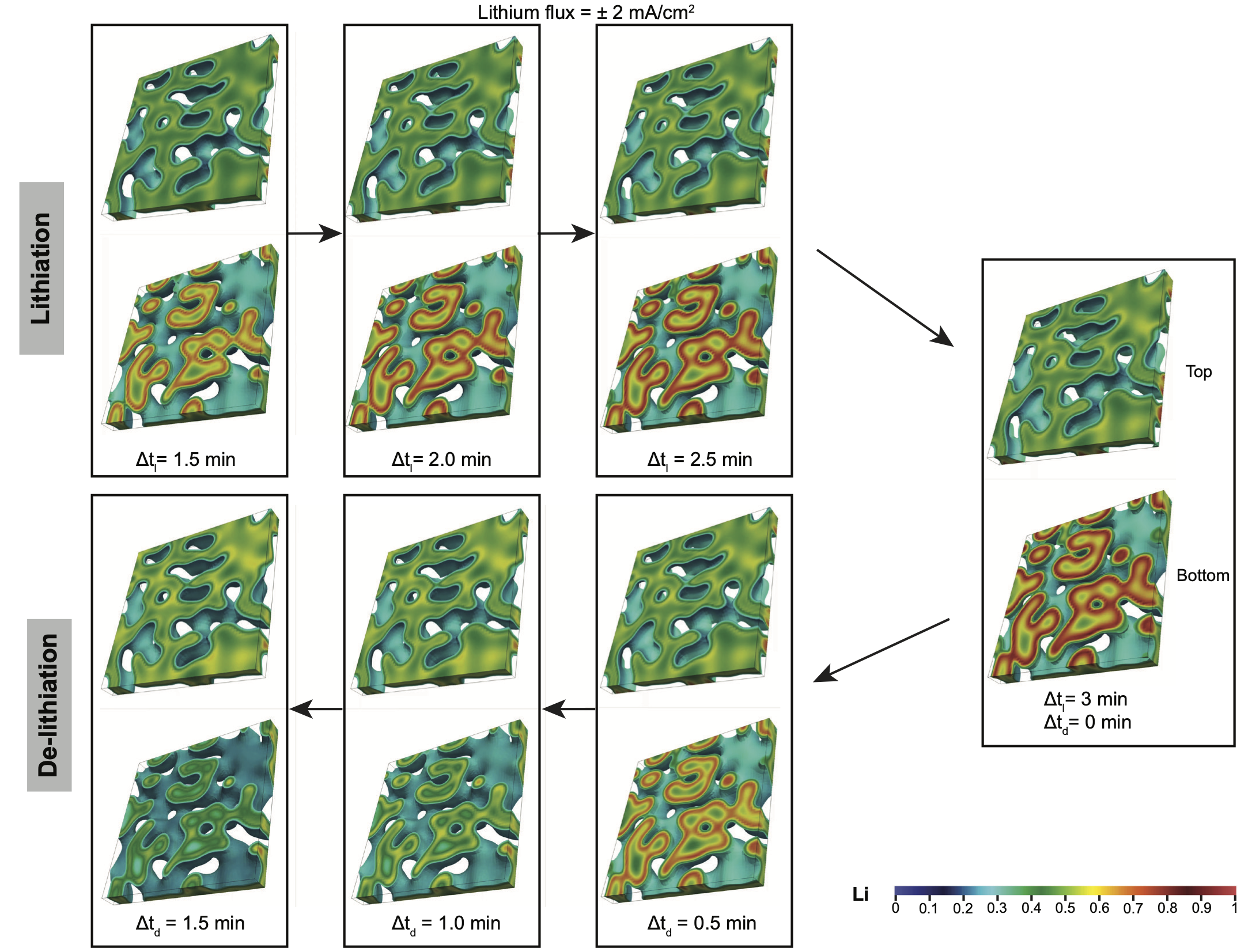}
    \caption{\footnotesize Morphogenesis during charge-discharge cycling of anode interface at Li flux of $\pm 2$ $mA/cm^2$. Starting from the top-left microstructure depicting the lithium composition on the top and bottom surface of the intermetallic, the top row presents the microstructure at instants during charging. The microstructure produced after 3 minutes then undergoes discharging to evolve into the microstructures in the bottom row. The microstructure in each column has the same amount of net lithium. }
    \label{fig:cyclinga}
\end{figure}

\begin{figure}[h!]
    \centering
       \includegraphics[width=0.75\linewidth]{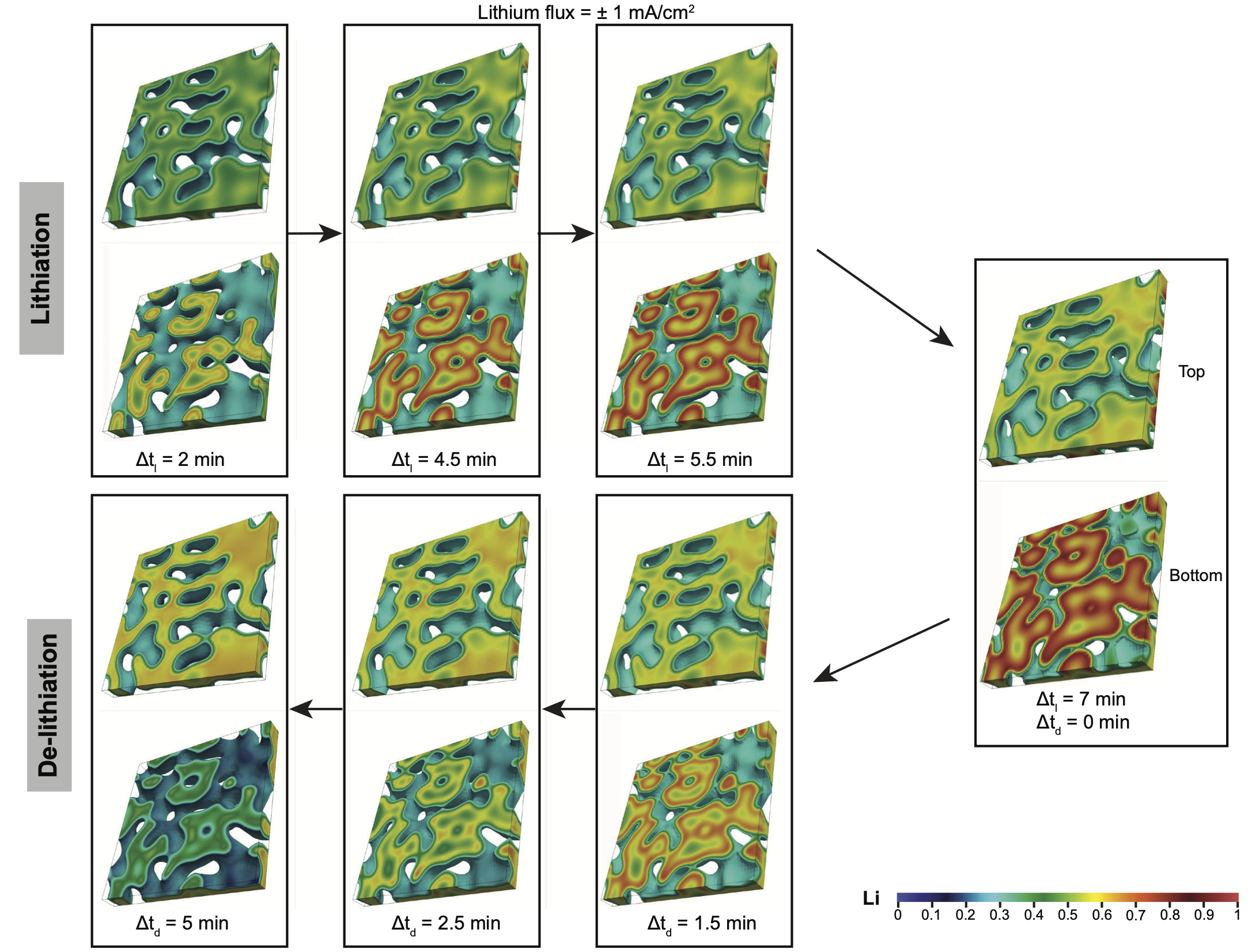}
    \caption{\footnotesize Morphogenesis during charge-discharge cycling of anode interface at Li flux of $\pm 1$ $mA/cm^2$.}
    \label{fig:cyclingb}
\end{figure}

\begin{figure}[h!]
    \centering
    \includegraphics[width=0.75\linewidth]{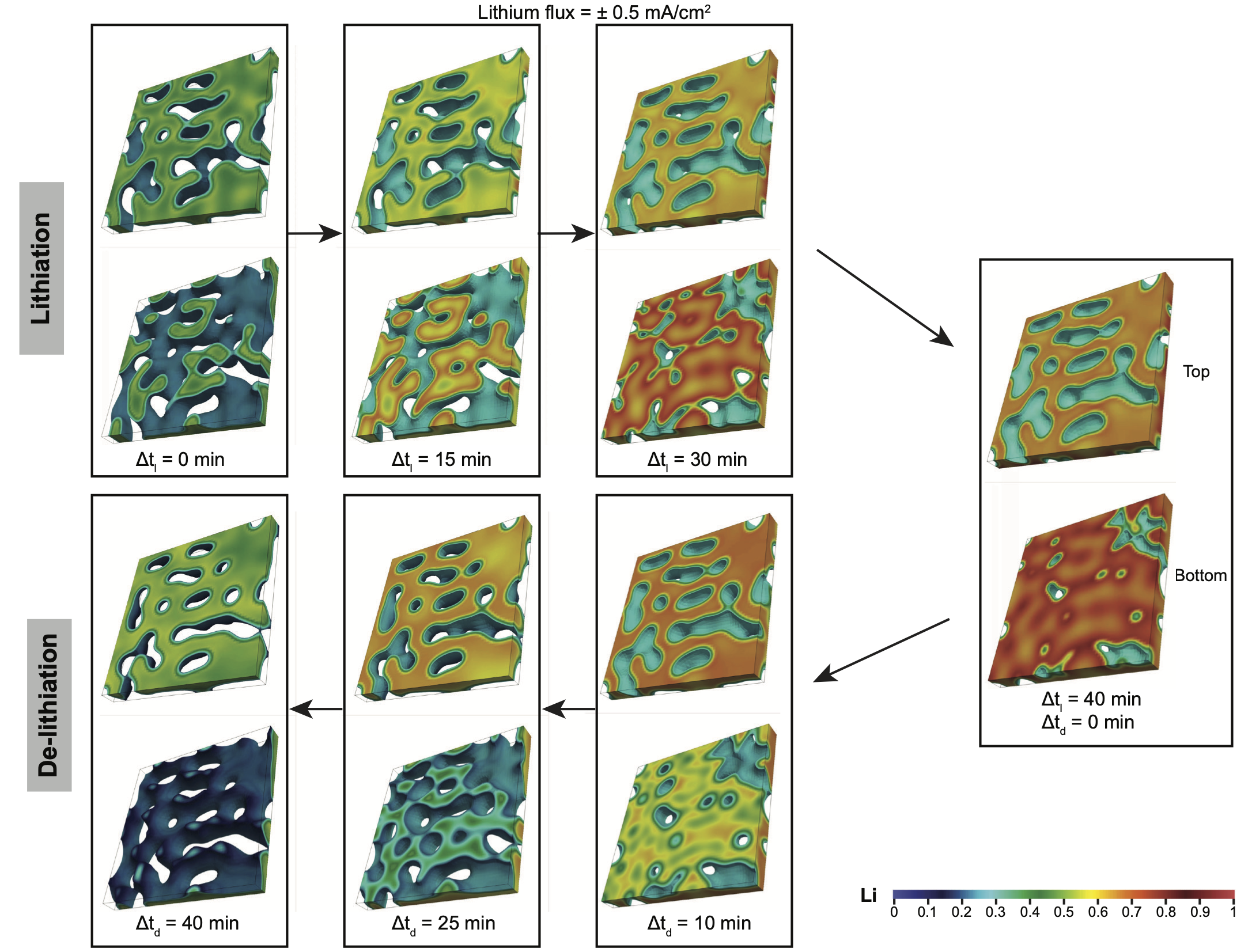}
    \caption{\footnotesize Morphogenesis during charge-discharge cycling of anode interface at Li flux of $\pm 0.5$ $mA/cm^2$.}
    \label{fig:cyclingc}
\end{figure}

\subsubsection{Charge discharge}

We next explore morphogenesis of the intermetallic layer after complete charge-discharge cycles with different fluxes imposed at Li mass-equivalent exchange times as in Figures \ref{fig:cyclinga}, \ref{fig:cyclingb} and \ref{fig:cyclingc}. The lithium flux in the respective figures is $\pm 2$ $mA/cm^2$, $\pm 1$ $mA/cm^2$ and $\pm 0.5$ $mA/cm^2$ respectively. In each of the cases, the initial pore structure is subjected to lithiation followed by de-lithiation. Here, $\Delta t_l$ represents the lithiation time and $ \Delta t_d$ is the delithiation time. The intermetallic layer is compared at the intermediate steps during charging and discharging respectively. We observe that the charge-discharge cycle leads to a different microstructure compared to the initial state, after the deposited lithium is completely extracted. Also note the characteristics of void formation at the contact surface over several charge-discharge-resting cycles, and the growth of lithium at the bottom and top surfaces during lithiation and depletion during delithiation. Such  explorations reveal how the microstructure of the battery  deviates with every added cycle, and could aid in optimizing the morphogenic structure of solid-state batteries.

\begin{figure}[h!]
    \centering
    \includegraphics[width=0.83\linewidth]{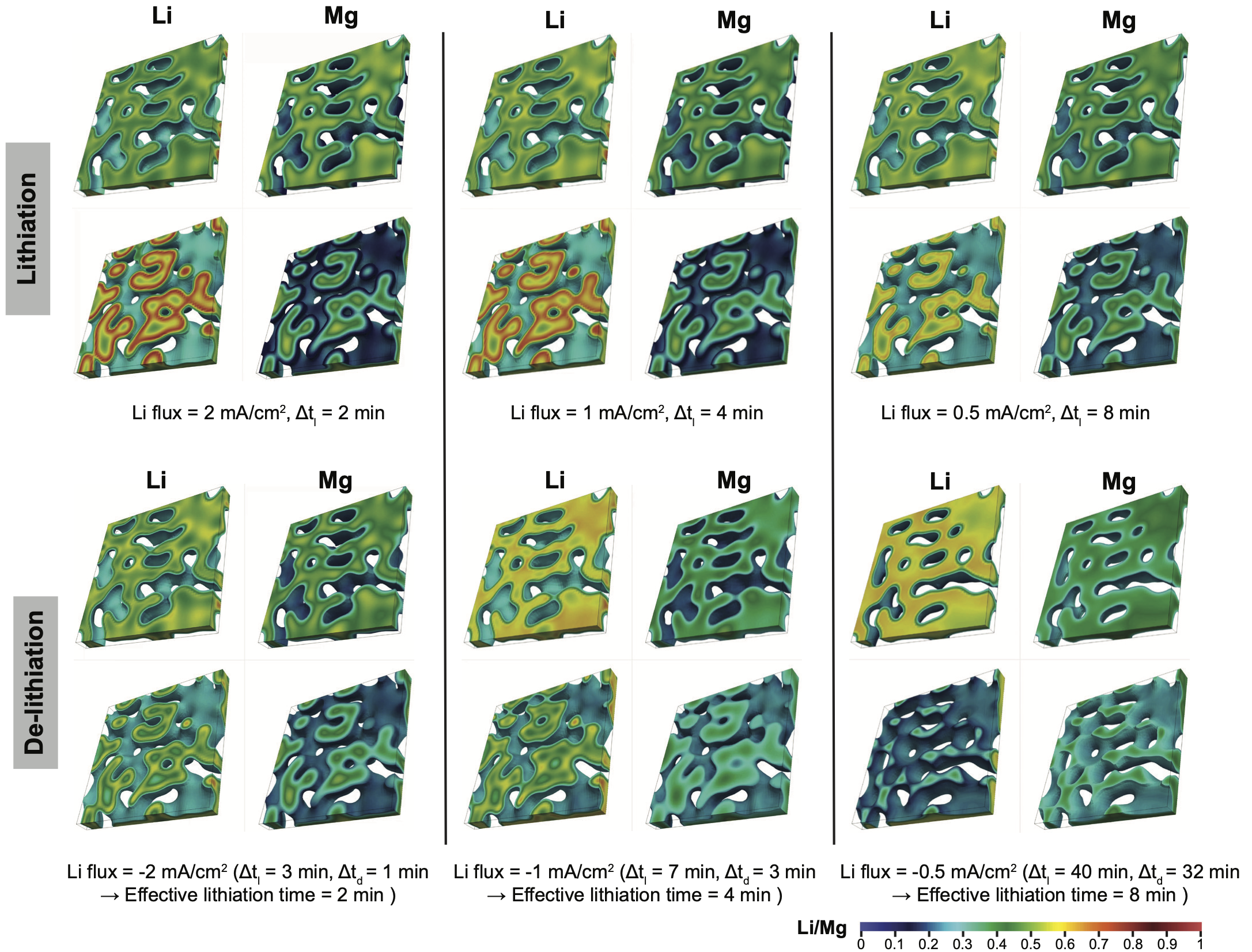}
    \caption{\footnotesize Snapshots of various fluxes applied during equivalent Li-exchange times in a cycle, with each case containing the same amount of Li/Mg but different morphologies and metal distributions. Starting from identical initial configurations, the snapshots in the first two rows, from left to right, display the lithiated interface resulting from fluxes of 2 $\text{mA}/\text{cm}^2$ for 2 minutes, 1 $\text{mA}/\text{cm}^2$ for 4 minutes, and 0.5 $\text{mA}/\text{cm}^2$ for 8 minutes, respectively. Consequently, each configuration contains the same amount of lithium but exhibits distinct morphology and metal distributions. The delithiation snapshots in the last two rows are obtained by first lithiating the interface for an extended period and then reversing the flux. For instance, with the 1 $\text{mA}/\text{cm}^2$ flux, the interface is lithiated for an additional 3 minutes (7 minutes total) and then delithiated for 3 minutes, reverting to the configuration at the 4-minute mark with the same amount of Li/Mg content but different morphology and Li/Mg distribution.}
    \label{fig:cyclying2}
\end{figure}

\begin{figure}[h!]
    \centering
    \includegraphics[width=.75\linewidth]{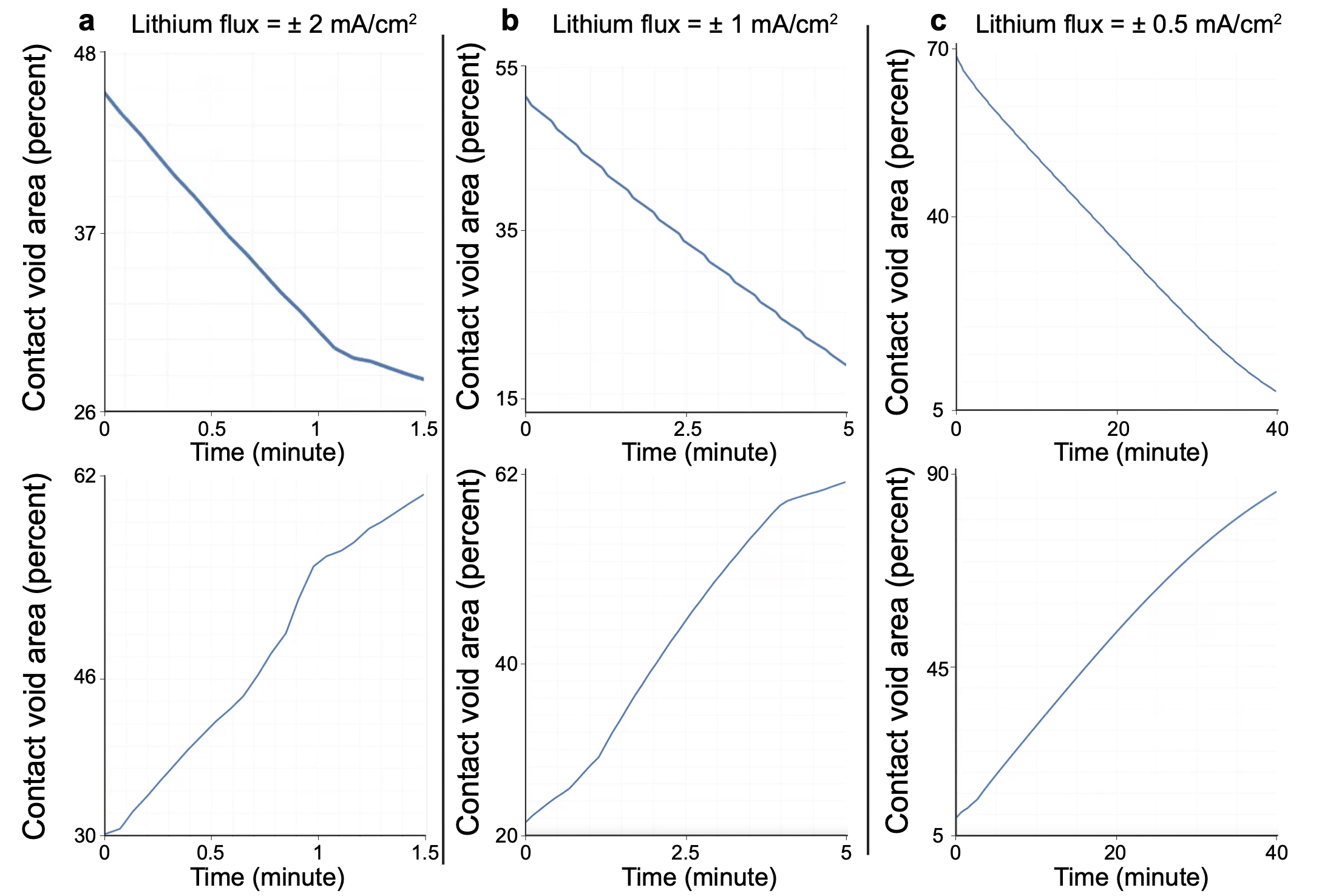}
    \caption{\footnotesize The percent void area on the contact surface during charging and discharging of the battery. }
    \label{fig:cyclying3}
\end{figure}

Figure \ref{fig:cyclying2} compares the resulting intermetallic microstructure after charging beyond a given time and then discharging the added lithium to return to the equal net lithium content.  The snapshots in the same column have the same amount of Li/Mg, as do the rows, but one can notice the decrease in Li content on the flux surface from left to right. This is due to lithium plating which is the result of higher lithium flux/fast charging of the anode. Furthermore, as in the previous case study, de-lithiation/discharging leads to a different microstructure  compared to the initial state. The images in the first column  represent the morphological evolution of the interface subjected to a lithium flux of $2$ $\text{mA/cm}^2$, after a lithiation time of $\Delta t_l = 2$ minutes. The interface is further subjected to lithiation for a total $\Delta t_l = 3$ minutes. At this stage, a delithiation flux is applied for $\Delta t_d = 1$ minute to obtain the comparable morphology with a net effective lithiation time of 2 minutes as shown in the bottom part of the same column. Similarly, the second column depicts the morphology for a lithium flux of $1 \text{mA/cm}^2$ after $\Delta t_l = 4$ minutes, and $\Delta t_d = 3$ minutes after starting from a total $\Delta t_l = 7$ minutes to arrive at the same net lithium concentration. The third column can be similarly understood.

The void area of the contact surface exchanging Li flux during the charging/discharging process is presented in Figure \ref{fig:cyclying3} for the respective cases of lithiation/delithiation fluxes imposed for the same durations, and therefore corresponding to the same amount of lithium exchanged in and out. Each column, however has different fluxes and durations of application. The decrease in contact void area after the net Li charge of 3 $mA.min/cm^2$ during charging for the three cases is approximately 18\%, 21\%, and 9\%, respectively. The increase in contact void area during discharging the same net Li value is 31\%, 28\%, and 9\% percent respectively. These comparisons again demonstrate that discharging at higher rates delithates the outflux surface to a greater extent than the interior of the layer. Also note the decreasing difference between changes in contact void areas as the charge/discharge rates decrease. Near-reversibility is obtained at the lowest rate of $\pm 0.5 \text{mA/cm}^2$. The variation in the contact area for case c), which has the lowest flux rate, is the smallest because morphogenesis involves a longer time duration here, allowing the added lithium to diffuse into the bulk interface during charging or allowing the diffusion of lithium from the bulk interface to the contact surface during discharging. Further, the change in contact void area for the three cases is not monotonic due to the different initial configurations. This study provides some insight to attaining lower void areas at the contact surface, which are desired for the longevity of a battery. 

\subsection{Morphogenesis in Li-Zn interface layers}
Lithium-zinc intermetallics are of interest, primarily for the high Li mobility these alloys. To set the stage we present the free energy density function motivated by DFT studies.

\begin{figure}[h!]
    \centering
    \includegraphics[width=0.7\linewidth]{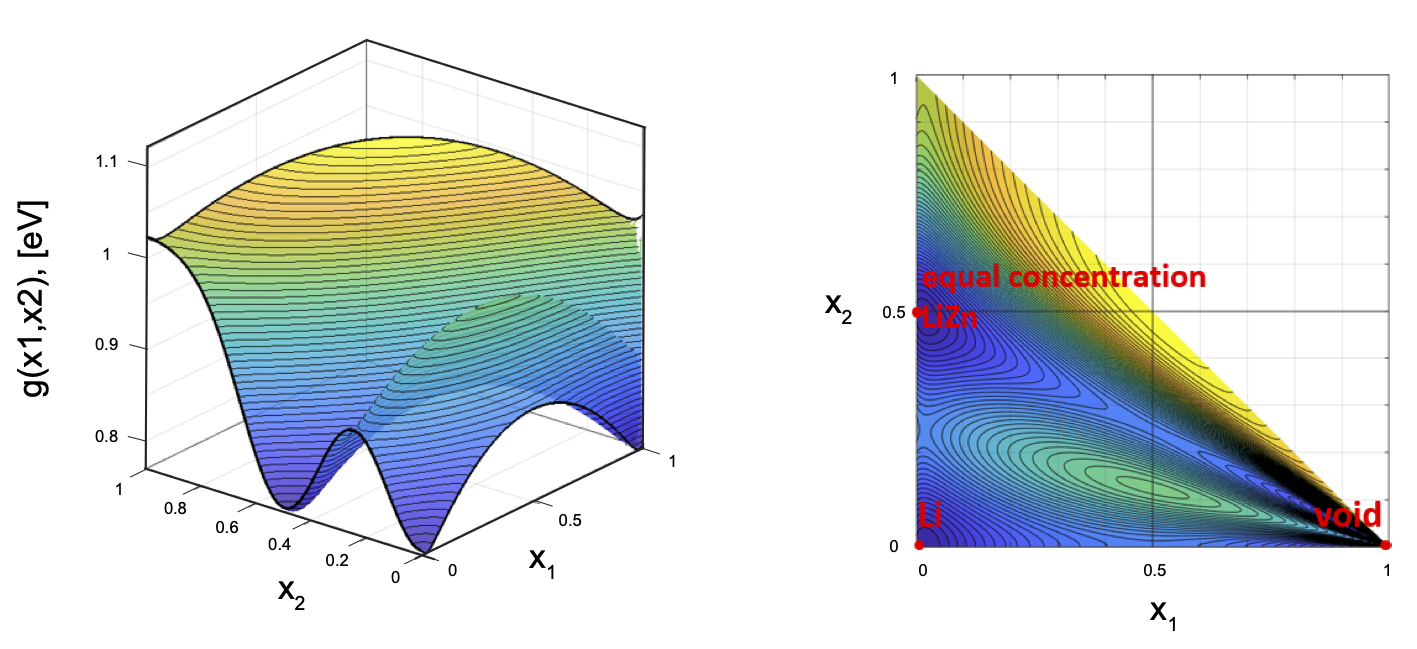}
    \caption{\footnotesize Free energy for Li-Zn alloys with vacancies (Eq. (\ref{eq:Li_Zn_enr})) depicting the location of the energy wells. The plot on the right emphasizes the constraint $x_1+x_2 \le 1$, implying $x_3 \ge 0$. } 
    \label{fig:Li_Zn_vac_enr}
\end{figure}

\subsubsection{DFT-informed free energy of Li-Zn alloy with vacancies}
We define the free energy for a Li-Zn alloy with vacancies by modifying our  model for Li-Mg alloys with vacancies in Eq. (\ref{eq:Li_Mg_enr}). In particular, we replace the original quadratic function with a Gaussian mixture model along the $x_2$ axis, which in the new model represents Zn composition in a ternary Li-Zn-vacancy alloy. As in the Li-Mg model, vacancies remain on the Li sub-lattice with composition $x_1$. The free energy is written as:
\begin{equation}\label{eq:Li_Zn_enr}
g(x_1,x_2) = g_1(x_1) +p\left[\exp\left(-q\left(\frac{x_2}{1 - x_1}\right)^2\right) + \exp\left(-q\left(\frac{x_2}{1 - x_1} - 0.5\right)^2\right)\right],
\end{equation}
where $p = 0.2511 \, \text{eV}$ and $q = 20$ is a dimensionless parameter. The modification introduces two distinct energy wells along the $x_2$ axis: the first at $(x_1, x_2, x_3) = (0+\varepsilon, 0, 1-\varepsilon)$ ($0 < \varepsilon \ll 1$), representing a pure Li phase, and the second at $(x_1, x_2, x_3) = (0+\varepsilon, 0.5, 0.5-\varepsilon)$, representing an equimolar Li-Zn phase. The  lithium-rich phase corresponds to $\beta$-Li. While the equimolar Li-Zn phase is succeeded by others at increasing Zn stoichiometry (some of them line compounds), there is no essentially important phenomenology that is revealed by modelling all of the high-zinc phases. We therefore have chosen to model only Li-Zn ($x_2 = 0.5, x_1 \to 0^+$). We also retain the energy well corresponding to voids at $(x_1, x_2, x_3) = (1-\varepsilon, 0, 0+\varepsilon)$. Increasing $q$ in Eq. (\ref{eq:Li_Zn_enr}) results in steeper wells along the $x_2$ axis. We optimized  $p$ and $q$ for the most accurate phenomenological responses. Figure \ref{fig:Li_Zn_vac_enr} illustrates the energy surface defined in Eq. (\ref{eq:Li_Zn_enr}).

\begin{figure}[h!]
    \centering
    \includegraphics[width=1\linewidth]{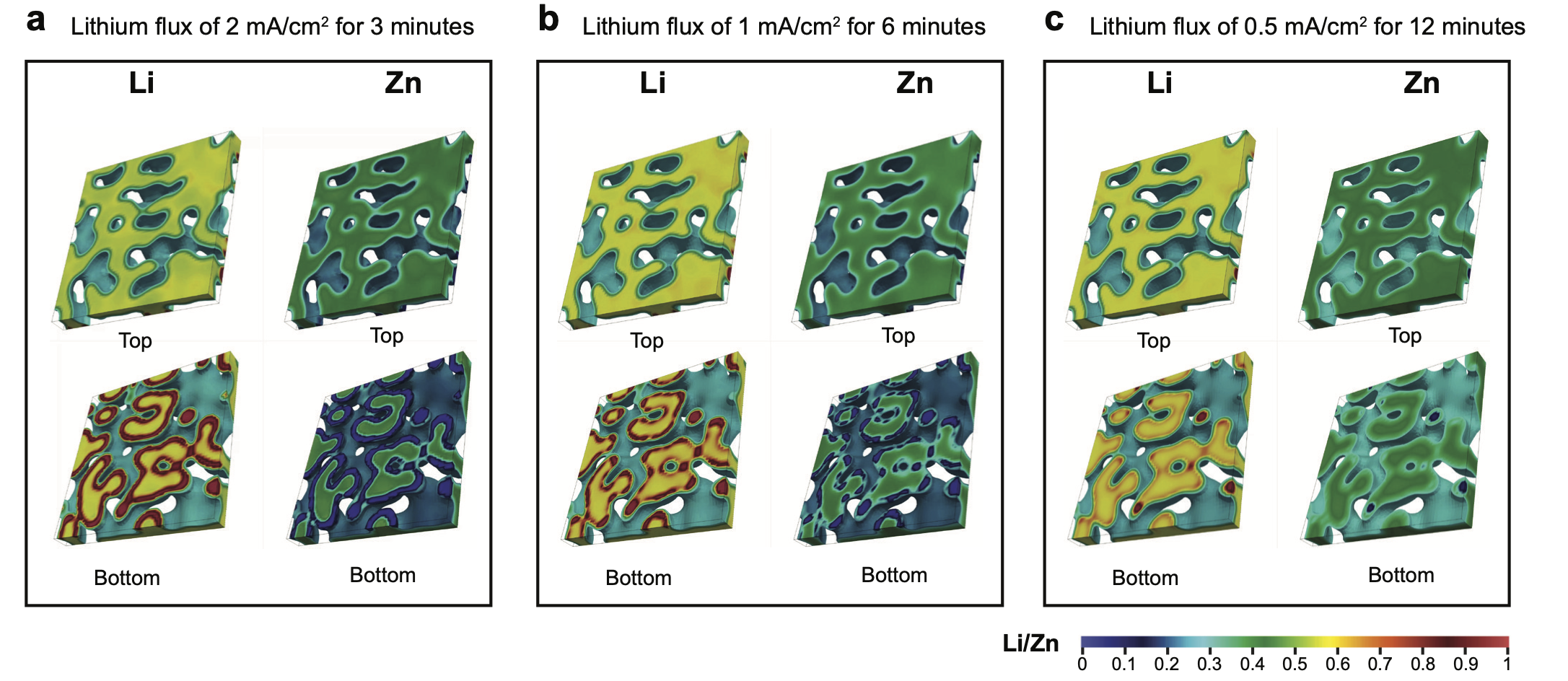}
    \caption{\footnotesize Li-Zn intermetallic evolution during charging at different Li fluxes but equal net Li transport. Note the formation of sharply delineated Li islands compared with the smoothly varying Li composition in Li-Mg (Figure \ref{fig:flux}). Slightly more lithium plating is observed at higher fluxes.}
    \label{fig:LiZn}
\end{figure}

\subsubsection{Charge cycling}
The parametric studies carried out for Li-Mg remain relevant for understanding the phenomenology of Li-Zn interface layer morphogenesis. Additionally, the phase separation that can occur along the Li-Zn axis drives some aspects of layer morphogenesis. We study it for charge cycling under different fluxes transporting the same net amount of lithium into and out of a Li-Zn porous layer (Figure \ref{fig:LiZn}). The new model effectively resolves the 3 phases; voids, Li-Zn and Li-rich $\beta$. Figure \ref{fig:LiZn} bears comparison with Figure \ref{fig:flux}, which illustrates a  study of the same fluxes and times for Li-Mg. We call attention to the sharper delineation of the high lithium regions for Li-Zn at all flux levels. This morphogenic feature is a consequence of the sharp segregation into the Li-rich $\beta$ phase and the equimolar Li-Zn phase, in contrast with the wider range of solubility of Li-Mg in the $\beta$-Mg phase which results in smoothly varying composition between Li-rich and Mg-rich regions in Figure \ref{fig:flux}. The surviving Li-rich islands in the LiZn case continue to provide pathways for vacancy transport. A slightly more pronounced lithium plating at the contact surface is observed at a higher flux rate as in the Li-Mg case study.

\subsection{Role of grain boundaries in morphogenesis}

Grain boundaries are vacancy sources and sinks. Because of the role that vacancies play in lithium transport we explored how this source/sink mechanism may alter interface layer morphogenesis. 

\subsubsection{A vacancy transport model with grain boundaries}
 We modify the Cahn-Hilliard equation for the vacancy composition $x_1$, presented in Eq. (\ref{eq:CH}), by adding a source term as follows:
\begin{equation}
    \frac{\partial x_1}{\partial t} = - \nabla \cdot (-M_1 \nabla \mu_1) -\frac{\delta_{GB}}{\tau} (x_1 - x_1^\text{eq}), \quad
    \delta_{GB} = 
    \begin{cases} 
        1 & \text{near grain boundaries} \\
        0 & \text{otherwise}
    \end{cases}.
    \label{eq:gbmodel}
\end{equation}
With $\delta_{GB}$ as an indicator function, the grain boundary locations within the material can be specified. Here, $x_1^{eq}$ denotes the equilibrium vacancy composition, and $\tau$ is the relaxation time. The grain boundary  source/sink mechanism drives the local composition $x_1$ toward $x_1^\text{eq}$ at a rate $1/\tau$.  This leads to a modified version of Eq. (\ref{eq:weakform}), which we solve for  zinc flux $j_{n_2} = 0$ and with  the same free energy function as detailed in the previous section for Li-Zn intermetallic systems:
\begin{equation}\label{eq:weakform_gb}
\begin{aligned}
    0 &= \int_V \left[ w_{x_1}\frac{\partial x_1}{\partial t} + M_1\nabla w_{x_1}\cdot\nabla{\mu}_1 + w_{x_1}\frac{\delta_{GB}}{\tau} (x_1 - x_1^{eq}) \right]\mathrm{d}v + \int_{\partial V} w_{x_1}j_{n_1}\mathrm{d}s\\
    0 &= \int_V \left[w_{\mu_1}\left(\mu_1 - \frac{\partial g}{\partial x_1}  \right) - \kappa_1\nabla w_{\mu_1}\cdot\nabla x_1\right]\mathrm{d}v\\  
    0 &= \int_V \left[ w_{x_2}\frac{\partial x_2}{\partial t} + M_2\nabla w_{x_2}\cdot\nabla{\mu}_2\right]\mathrm{d}v \\
    0 &= \int_V \left[w_{\mu_2}\left( \mu_2 - \frac{\partial g}{\partial x_2}  \right) - \kappa_2\nabla w_{\mu_2}\cdot\nabla x_2\right]\mathrm{d}v.\\  
\end{aligned}
\end{equation}

\subsubsection{Grain boundaries as  diffusion pathways}
\label{sec:gb0}
\begin{figure}[h!]
    \centering
    \includegraphics[width=0.7\linewidth]{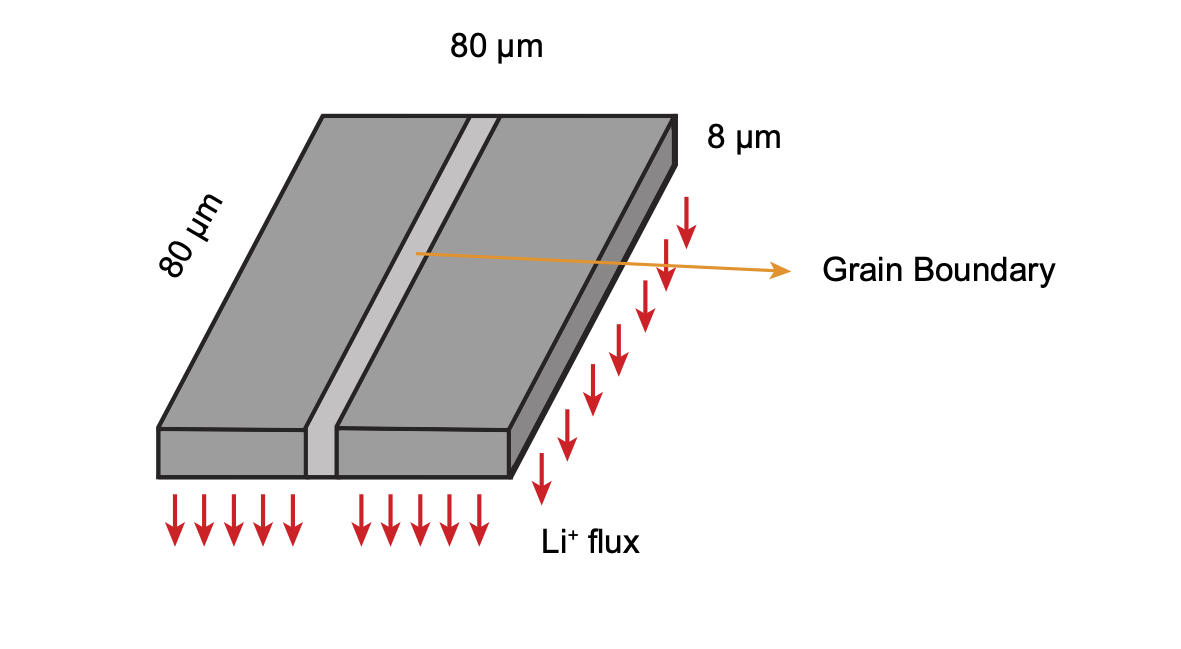}
    \caption{ Schematics of the Li-Zn interlayer containing the grain boundary used for the numerical simulation.}
    \label{fig:schematicsgrainBoundary}
\end{figure}

\begin{figure}[h!]
    \centering
    \includegraphics[width=1\linewidth]{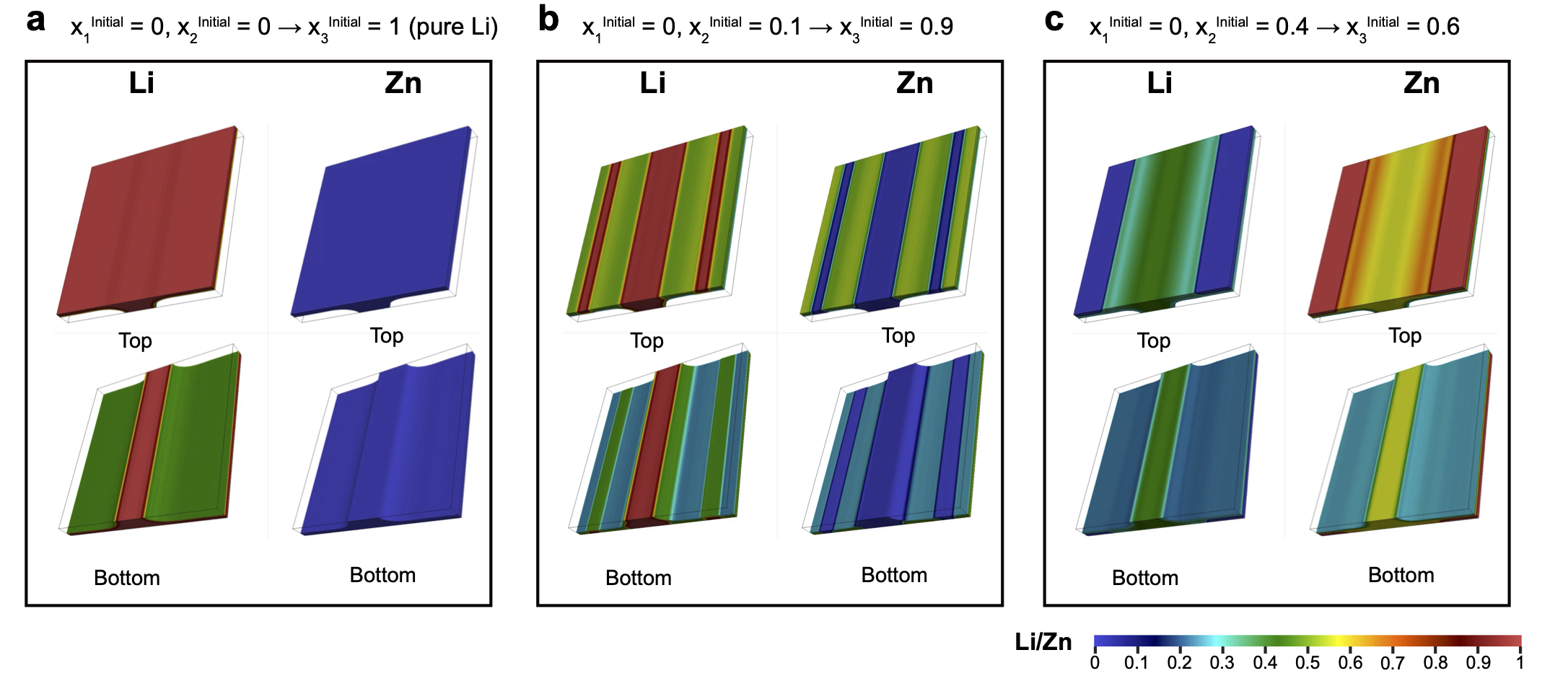}
    \caption{ Role of grain boundary as a source of vacancies and in preserving metal along the interface. The grain boundary runs along the entire length of the portion of the intermetallic layer being modeled. Cases (a), (b), (c) show  morphogenesis for increasing amounts of initial Zn ($x_2$). Note Li segregation in the neighborhood of the grain boundary, and the development of the LiZn phase.}
    \label{fig:grainBoundary}
\end{figure}

As a baseline study of its role as a vacancy source and sink, we model a grain boundary with a de-lithiation flux of $0.5$ $mA/cm^2$ at the contact surface as illustrated in Figure \ref{fig:grainBoundary}a-c. The schematics of the grain boundary and the vacancy influx surface are depicted in Figure \ref{fig:schematicsgrainBoundary}. In the studies here, the region over which $\delta_{GB} = 1$ was chosen to be a bit wider than the sub-nanometer widths of physical grain boundaries, thus representing a smeared-out diffuse grain boundary model that saves the numerical expense of finite element mesh resolution, while mollifying the numerical stiffness of the computations.  Sharp interface  resolving finite element methods such as those in Ref \cite{zhang2024treatment} present be an alternate computational approach to the diffuse model employed here. The three cases in Figure \ref{fig:grainBoundary}a-c are for increasing initial Zn compositions. The Li outflux on the bottom surface (delithiation) is represented by a vacancy influx. Along the grain boundary, excess vacancies are consumed, which increases the Li composition, $x_3$ as these atoms undergo short-range transport from the neighborhood of the grain boundary. Effectively, this creates a  diffusion path for Li. As Li suffers depletion away from the grain boundary, it leads to the formation of the Li-Zn phase in cases (b) and (c), which have some initial Zn.

\subsubsection{Parameteric study of the strength of the source term}
\label{sec:gb1}
\begin{figure}[h!]
    \centering
    \includegraphics[width=1\linewidth]{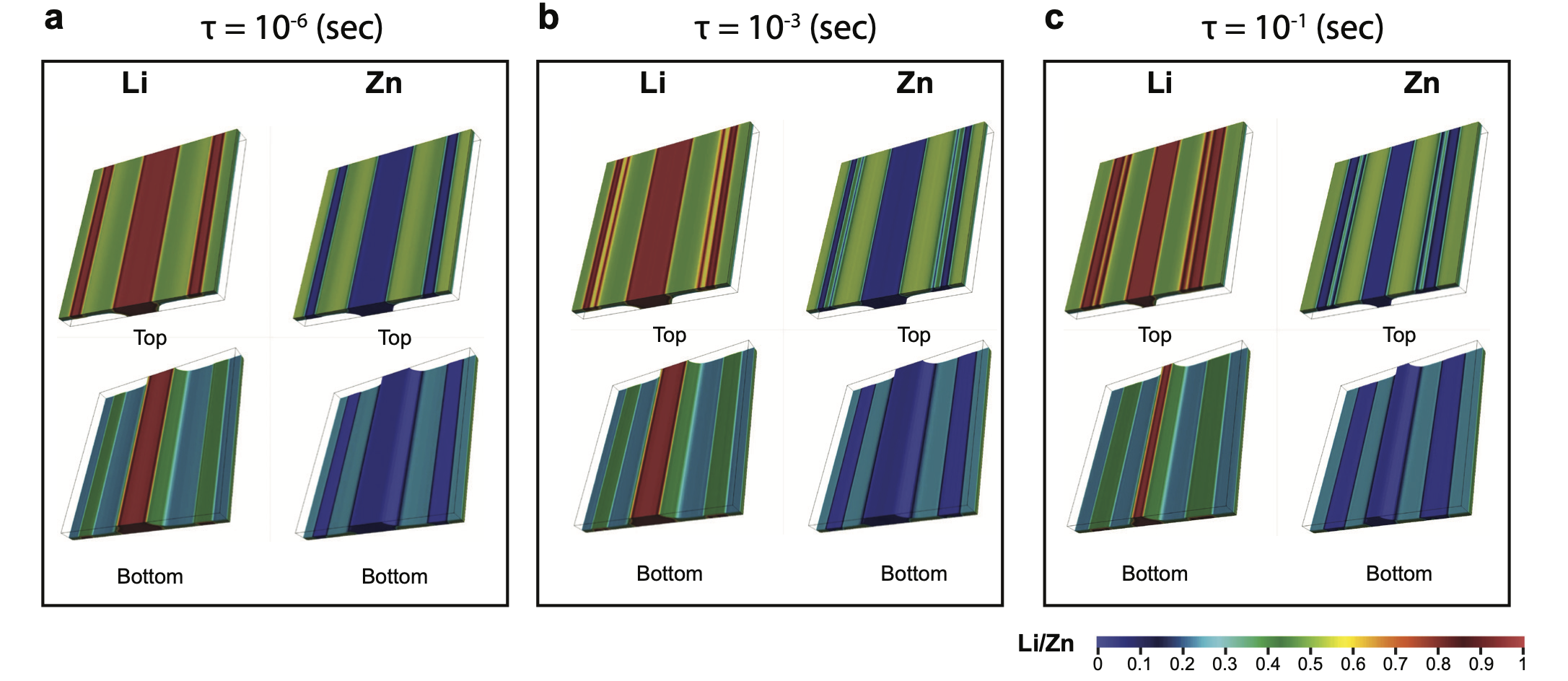}
    \caption{\footnotesize Effect of the relaxation time $\tau$ on the morphogenesis of the  diffusion path for Li. As $\tau$ increases, the diffusion pathway becomes degraded.}
    \label{fig:grainBoundarySource}
\end{figure}

The role of the relaxation time $\tau$ is explored in Figure \ref{fig:grainBoundarySource}. 
Increasing $\tau$ corresponds to slower relaxation of $x_1$ toward $x_1^\text{eq}$. This may also be understood as $\tau^{-1}$ acting as a penalty term on the vacancy composition deviating from its equilibrium value in Eq. (\ref{eq:gbmodel}).  As $\tau$ is increased, this penalty is relaxed, vacancy formation/consumption slows down, and with it Li segregation in the grain boundary is weakened. Effectively, increasing $\tau$ leads to a less prominent vacancy source/sink mechanism at the grain boundary and the strength of the diffusion path for Li is degraded.

\subsubsection{Triple junction in the porous interface layer }

\begin{figure}[h!]
    \centering
    \includegraphics[width=1\linewidth]{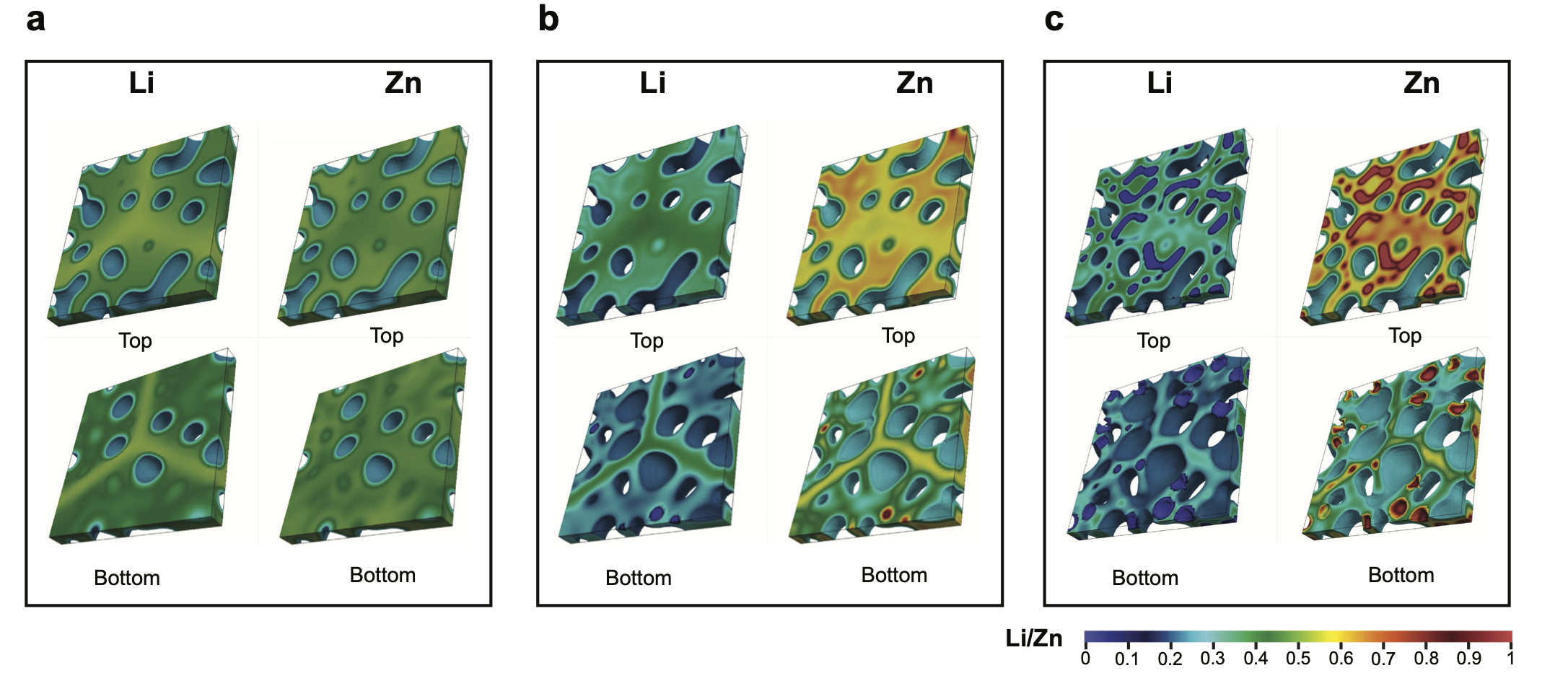}
    \caption{\footnotesize The width of the region over which the grain boundary source/sink mechanism acts has a pronounced role in the evolution of the intermetallic microstructure.  This width $\delta_{GB}$ decreases by a factor of $1/2$ from (a) to (c). }
    \label{fig:grainBoundary2}
\end{figure}

Finally, the above study was extended to a triple junction grain boundary representing a more realistic microstructure (Figure \ref{fig:grainBoundary2}). We start with an initial Li-Zn porous structure subjected to a de-lithiation flux of $0.5$ $mA/cm^2$. In this case the width over which the vacancy source/sink acts is varied, decreasing by a factor of $1/2$ from Figure \ref{fig:grainBoundary2}a to  \ref{fig:grainBoundary2}c. The same physical mechanisms delineated in Sections \ref{sec:gb0} and \ref{sec:gb1} play out, in the more realistic triple junction microstructure. The grain boundaries develop into  diffusion paths for Li. The triple junction  ensures that this mechanism is accessible to more of the interface layer. The effect on morphogenesis of the interface layer is to cause more rapid depletion of Li from the entire layer. The existing voids grow and new ones form. There is extensive void formation as $\delta_{GB}$ decreases, the vacancy consumption mechanism weakens and the non-absorbed vacancies coalesce into voids. The loss of Li gives rise to the equimolar LiZn phase at smaller values of $\delta_{GB}$.

\section{Conclusion}\label{sec:conclusion}

We have carried out computational studies of the morphogenesis of  lithium intermetallic layers, motivated by experimental studies on anodes. Charge-discharge cycling  leads to microstructural changes: not only does the distribution of lithium and the alloying metal evolve, but the porous structure of the interface layer also undergoes dynamic changes. Plating versus deep alloying of lithium, segregation versus smooth mixing and extensive void formation leading to contact area losses are just some of the effects that in turn will influence the capacity and cyclability of cells. 

We have nominally assumed a solid electrolyte, for which case charge transport across the contact layer can be affected by void fraction and surface lithium distributions. For instance,  our studies of lithium plating/alloying  can provide initial conditions for dendrite growth. Our computations are at the continuum scale using phase field methods. The free energy functions for the lithium-vacancy system  have been informed by DFT calculations, which we have reparameterized slightly in order to mollify the numerical stiffness that arises from pure metal and void phases existing close to $0$ and $1$ in composition space. The extensions accounting for specific intermetallics (Li-Mg and Li-Zn) have been guided by experimental phase diagrams. In this communication we have not included mechanics, chiefly because extensive parameterization of chemo-mechanical coupling would be needed. However, the voided interface layers arising in the computations presented here could be subjected to compressive loading to determine the external pressure needed to maintain contact. Coupled chemo-mechanical studies would include this as a dynamic effect during charge/discharge cycles. For recent work on continuum modelling of chemo-mechanically driven failure we point to Ref. \cite{zhang2024treatment}, and to Ref. \cite{van2023ferroelastic} for a perspective on possible toughening mechanisms in solid state batteries. 

The kinetics of lithium/metal transport is another complex, composition- (and stress-) dependent phenomenon. Experimental and first principles computational results for the same system differ by several orders of magnitude. We have therefore used baseline values, which we have then subjected to parameter sweeps to study the influence of mobility. Its competition with the charge rate ((de)lithiation flux) leads to the lithium-metal segregation/smooth distribution discussed above, as well as to the void-driven morphogenesis. Irreversibility was seen to emerge even over short charge/discharge times. Only at slower charge/discharge rates does the interface layer morphology return closer to its initial state. The exploration of this aspect of morphogenesis is relevant to cyclability and capacity fade. 

We have modelled Li-Mg, which has a wide solubility range in the $\beta$-Mg phase, and Li-Zn, which has several phases at Zn compositions exceeding 1/2. While the broad morphogenic effects discussed above are common, there are some differences, mainly in the distribution of Li and Zn, which can be more sharply delineated than Li and Mg. These distributions have implications for maintenance of metal contact. Our phase field treatment exploits the role of vacancies, which we have modelled as undergoing transport on the lithium sub-lattice. Therefore, lithiation/delithiation is represented as vacancy outflux/influx. From this perspective, it has been natural to study the role of grain boundaries, given the vacancy source/sink mechanism that they control. Our studies show that grain boundaries that have a strong source/sink mechanism produce diffusion paths for lithium, providing a conduit for current even at high degrees of delithiation. 

In summary, this work contributes a novel perspective on the morphogenesis of intermetallic anode-electrolyte interfaces with relevance to solid-state batteries, with a focus on the three-dimensional morphogenic evolution of the  interface.
The framework presented here can be a stepping stone to further studies including: (a) exploiting interface morphogenesis to optimize current charge density and cyclability, (b) optimizing the selection of lithium intermetallic based on the desired lithium mobility, (c) tailoring the anode grain structure to leverage the availability of vacancy sources/sinks, (d) coupling with mechanics, both at the atomic scale via free energy functions that resolve local chemo-mechanical interactions, and at the continuum scale for interface void formation and contact maintenance. Experimental data on diffusivity has been used to a small degree in our study. As a step toward first principles DFT and statistical mechanics-informed studies, we call attention to Refs. \cite{behara2024fundamental,thomas2024thermodynamic}. A combined campaign using first principles and phenomenological continuum computations and  supported by  extensive experimental data, including imaging of interface layers as a function of charge-discharge cycles, will advance the notion of morphogenesis as a design approach for solid state batteries. 

\section*{Acknowledgements}\label{sec:acknowledgement}
This work was supported by GE Vernova via a direct grant  \#HR001122C0097 from DARPA under the Morphogenic Interfaces (MINT) program. The authors wish to acknowledge many useful discussions with Sesha Sai Behara, Anton Van der Ven and Joseph Shiang.

\section*{Code Availability}
The source code used in this work is available as an open source library \cite{GitRepo2024}.
\small
\bibliographystyle{unsrt}
\bibliography{main}

\end{document}